\RequirePackage{fix-cm}
\documentclass[smallextended,natbib]{svjour3}       
\smartqed  
\usepackage[bookmarksdepth=2]{hyperref}
\usepackage[dvipsnames]{xcolor}
\usepackage{colortbl} 
\usepackage{graphicx}
\usepackage{subfigure}
\usepackage{xspace}
\usepackage{listings}
\usepackage{tabularx}
\usepackage{tablefootnote}
\usepackage{booktabs}
\usepackage{multirow}
\usepackage{enumitem}
\usepackage[most]{tcolorbox}
\usepackage[T1]{fontenc} 
\usepackage{pythonhighlight}
\usepackage{float} 
\usepackage{url}
\usepackage{makecell}

\definecolor{summary_box}{RGB}{128,128,128}

\newtcbtheorem{Summary}{\bfseries Summary of Research Question}{enhanced,drop shadow={black!5!white},
  coltitle=white,
  top=0.3in,
  attach boxed title to top left=
  {xshift=1.5em,yshift=-\tcboxedtitleheight/2},
  boxed title style={size=small,colback=summary_box}
}{summary}

\usepackage{color}
\definecolor{deepblue}{rgb}{0,0,0.5}
\definecolor{deepred}{rgb}{0.6,0,0}
\definecolor{maroon}{rgb}{0.5,0,0}
\definecolor{deepgreen}{rgb}{0,0.5,0}

\lstdefinelanguage{XML}
{
  basicstyle=\ttfamily,
  morestring=[s]{"}{"},
  morecomment=[s]{?}{?},
  morecomment=[s]{!--}{--},
  commentstyle=\color{darkgreen},
  moredelim=[s][\color{black}]{>}{<},
  moredelim=[s][\color{red}]{\ }{=},
  stringstyle=\color{blue},
  identifierstyle=\color{maroon}
  linewidth=.99\textwidth,
  frame=trbl,
  rulecolor=\color{black!40},
  backgroundcolor=\color{white},
  frame=single,breaklines=true%
}

\begin{document}

\title{Does Using Bazel Help Speed Up Continuous Integration Builds?	
}


\author{Shenyu Zheng \and
        Bram Adams \and Ahmed E. Hassan 
}


\institute{Shenyu Zheng \at
              School of Computing, Queen's University, Kingston, ON, Canada \\
              \email{22sz3@queensu.ca}           
           \and
           Bram Adams \at
              School of Computing, Queen's University, Kingston, ON, Canada \\
              \email{bram.adams@queensu.ca}
          \and
          Ahmed E. Hassan \at
          School of Computing, Queen's University, Kingston, ON, Canada \\
          \email{hassan@queensu.ca}
}

\date{Received: date / Accepted: date}

\maketitle

\begin{abstract}
A long continuous integration (CI) build forces developers to wait for CI feedback before starting subsequent development activities, leading to time wasted. In addition to a variety of build scheduling and test selection heuristics studied in the past, new artifact-based build technologies like Bazel have built-in support for advanced performance optimizations such as parallel build and incremental build (caching of build results). However, little is known about the extent to which new build technologies like Bazel deliver on their promised benefits, especially for long-build duration projects.

In this study, we collected 383 Bazel projects from GitHub, then studied their parallel and incremental build usage of Bazel in popular CI services (GitHub Actions, CircleCI, Travis CI, or Buildkite), and compared the results with Maven projects. We conducted 3,500 experiments on 383 Bazel projects and analyzed the build logs of a subset of 70 buildable projects to evaluate the performance impact of Bazel's parallel builds. Additionally, we performed 102,232 experiments on the 70 buildable projects' last 100 commits to evaluate Bazel's incremental build performance. Our results show that 31.23\% of Bazel projects adopt a CI service but do not use Bazel in the CI service, while for those who do use Bazel in CI, 27.76\% of them use other tools to facilitate Bazel's execution. Compared to sequential builds, the median speedups for long-build duration projects are 2.00x, 3.84x, 7.36x, and 12.80x, at parallelism degrees 2, 4, 8, and 16, respectively, even though, compared to a clean build, applying incremental build achieves a median speedup of 4.22x (with a build system tool-independent CI cache) and 4.71x (with a build system tool-specific cache) for long-build duration projects. Our results provide guidance for developers to improve the usage of Bazel in their projects, and emphasize the importance of exploring modern build systems due to the current lack of literature and their potential advantages within contemporary software practices such as cloud computing and microservice.

\keywords{Build Systems \and Continuous Integration \and Empirical Study}
\end{abstract}

\section{Introduction}
\label{sec:introduction}
Continuous Integration (CI) is the practice of frequently integrating software changes and checking the quality of the resulting integration through automated builds and testing. CI helps developers to increase productivity and maintain code quality \citep{vasilescu2015quality} \citep{hilton2016usage}, and is a widespread practice in the industry. An ideal CI build should take a short time to finish \citep{duvall2007continuous} since long builds delay the feedback, thus slowing down development activities \citep{hilton2017trade} \citep{zampetti2020empirical} \citep{bernardo2023cipull}.

However, recent studies show that, in practice, long CI builds are a pervasive phenomenon. While surveys on practitioners show that the most acceptable build duration is 10 minutes \citep{hilton2017trade}, \cite{ghaleb2019empirical}'s analysis of 104,442 CI builds from 67 GitHub projects found that 40\% of CI builds took more than 30 minutes to finish. Given the increasingly frequent delivery of software that is driven by industry practices such as continuous delivery \citep{humble2010continuous} \citep{chen2015continuous}, the negative impact of long build times will be more significant.

Many approaches have been proposed to deal with long CI builds such as optimizing CI scheduling by build outcome prediction (skipping builds for certain commits) or commit batching (building multiple commits together) \citep{jin2020cost} \citep{beheshtian2021software} \citep{divya2023pragmatic}, reducing CI testing time by test selection and prioritization \citep{pan2022test}, or speeding up builds by refactoring source code, e.g., removing redundant header files in C/C++ projects \citep{dayani2005improving} \citep{mcintosh2016identifying}. Practitioners have also invested extensive efforts in improving build technologies to enhance their performance.

Traditionally, features aimed at optimizing the performance of build technologies, such as parallel builds and incremental builds often come at the cost of correctness \citep{adams2016modern}. Traditional file-based (e.g., Make) or task-based technologies (e.g., Maven) usually lack the necessary information to generate an accurate dependency graph that determines the execution order of a build. This absence of dependency information can result in incorrect parallel and incremental builds \citep{licker2019detecting}, as well as non-reproducible builds \citep{lamb2021reproducible}. Moreover, even when these features successfully produce correct build results, their applicability within a CI context is often limited. For example, although incremental build is widely supported in major build technologies, they are rarely employed in CI due to the lack of support for remote caching \citep{maudoux2017bringing}. Developers often have to resort to external tools to share build outcomes between CI builds. It is noteworthy that while most CI services provide support for parallel CI builds, the CI build parallelism is generally achieved by scheduling entire jobs in parallel. The long compilation time of individual build jobs can still be a blocker of long CI builds no matter how many parallel jobs developers enable.

In recent years, a new family of artifact-based build system technologies such as Bazel \citep{bazeldoc}, Buck \citep{buckdoc}, and Pants \citep{pantsdoc}, has received a lot of attention, as it proposes a unique approach achieving both build performance and correctness, as well as built-in remote caching support. In 2015, Google released Bazel as an open-source version of its internal Blaze build technology. Unlike the traditional file-based or task-based build system technologies, Bazel's artifact-based philosophy enables developers to declaratively describe the artifacts and their dependencies in the build files. Based on this information, Bazel determines how to carry out the build. Because Bazel has precise knowledge about the build dependencies and has full control over the build process, it can generate an accurate dependency graph of the project. Therefore, Bazel provides reliable support of parallel and incremental builds, while ensuring the builds are correct and deterministic. As the build results are deterministic, Bazel allows developers to share build results across multiple machines through its remote caching to further improve performance and build reproducibility.

While the features of artifact-based build technologies like Bazel are promising, there is only limited research on the usage of artifact-based systems like Bazel within open-source projects. A lot of questions related to Bazel (and artifact-based build technologies in general) still remain to be answered, especially in the CI context. How do developers use Bazel in their projects? How useful are Bazel's features in a CI context? How much can Bazel speed up long-build duration projects? In fact, a recent study by \cite{alfadel2024icse} studied a phenomenon of organizations migrating away from build tools like Bazel due to higher build system maintenance costs than expected, for lower build performance gains. Instead, we are eager to understand the performance gains that realistically can be expected by Bazel’s parallelization and caching features, in order to help organizations determine whether Bazel (and its alternatives) would be worth the switch in the first place.

Therefore, we conducted an empirical study on 383 GitHub projects that use Bazel, focusing on the build system's compilation activities (similar to other work in this domain \citep{robles2006mining} \citep{maes2022revisiting} \citep{misu2024sourcererjbf}). From this sample, we selected projects that use GitHub Actions, CircleCI, Travis CI, or Buildkite as their CI services and examined their CI configuration files. We first examine whether developers use Bazel in CI services. Then we investigate the usage of parallel and incremental build features in CI services. We also make comparisons with Maven projects to gain a broader perspective. Moreover, we performed 3,500 experiments on the 383 projects' latest snapshots with 5 different parallelism degrees to assess the performance of the parallel build of Bazel. From the experiments, we identified 70 buildable projects divided into three groups based on their build duration, to evaluate the impact of parallel build performance on projects on projects with different build durations. Then, we analyzed the structure and granularity of the dependency graph of the projects to investigate their impact on the parallelization efficiency. Additionally, we conducted 102,232 experiments with 4 cache strategies on the last 100 commits of the 70 buildable projects to evaluate the incremental build of Bazel. As such, we address the following research questions.

\textbf{RQ1:} \textit{To what extent do developers use parallelization and incremental build (cache) features in CI builds?}

For the 4 studied CI services, we observed that 31.23\% of Bazel projects adopt a CI service but do not use Bazel in the CI service, as Bazel is adopted in these projects not only for building projects but also for other purposes such as making projects compatible to other Bazel projects (e.g., Non-Bazel C++ projects cannot be directly used as a library for Bazel C++ projects unless they also adopt Bazel in their projects). For projects that use Bazel in CI, we found that in CI builds, 26.36\% of them use tools such as Shell scripts, Make, and Docker to facilitate the execution of Bazel. All studied projects in the 4 studied CI services use Bazel's parallel build feature. However, surprisingly, while Bazel is famous for its good support of incremental build, we found only 44.08\% of them use caching to speed up build performance.

\textbf{RQ2:} \textit{What is the impact of Bazel parallelization on the build performance?}

We discovered that Bazel's parallelization can significantly (large effect sizes) improve the build performance at parallelism degrees 4, 8, and 16 for long-build duration projects. However, at parallelism degrees 2, short, medium, and long-build duration projects show no significant difference in speedups. We also found that, as the parallelism increases, the build performance tends to improve slowly for short and medium-build duration projects, while long-build duration projects still exhibit a continuous performance improvement. Nonetheless, at parallelism degree 16, 100\% of short, 96\% of medium, and 83\% of long-build duration projects are unable to fully harness the benefits of parallelism.

Moreover, unexpectedly, there is no significant correlation between the analyzed structural properties of the Bazel dependency graph with speedups, for any of the four parallelism degrees. However, the average size of the Bazel build targets in the dependency graph does show a significant correlation with build time speedups, implying that, at parallelism degree 16, there is a higher likelihood of improved build performance when the developers define smaller Bazel build targets for projects.

\textbf{RQ3:} \textit{What is the impact of Bazel incremental build (cache) functionality on the build performance?}

We found that, while caching of build dependencies is commonly employed in CI workflows, downloading dependencies from a remote cache in CI builds can still slow down the builds, with only 52.17\% of short, 8.70\% of medium, and 4.17\% of long-build duration projects exhibiting improved build performance compared to clean builds. 

We observed that caching of build results (i.e., incremental build) can greatly help reduce the build time for medium- and long-build duration projects. Specifically, for projects with medium build durations, we observed a median speedup of 1.21x when using a build system tool-independent CI cache ("\textit{General-Deps-and-Results} strategy") and a median speedup of 1.25x when using a build system tool-specific cache ("\textit{Specific-Deps-and-Results} strategy"). For projects with long build durations, the median speedups were even more notable, i.e., 4.22x and 4.71x, respectively. Nevertheless, the benefits of employing incremental builds for short-build duration projects are limited. In such cases, the differences in build speedups between the \textit{Specific-Deps-and-Results} and \textit{General-Deps} strategies, as well as between the \textit{General-Deps-and-Results} and \textit{General-Deps} are negligible and small, respectively.

Additionally, we discovered similar build performance improvement between \textit{General-Purpose CI Cache} and \textit{Build-Tool-Specific Cache} for incremental builds, as there is no significant difference in speedups between \textit{General-Deps-and-Results} and \textit{Specific-Deps-and-Results} strategies for medium- and long-build duration projects, while a significant difference with a small (0.179) effect size for short-build duration projects.

The paper is structured as follows. Section~\ref{sec:background} introduces the concepts of task-based and artifact-based build systems, as well as the related work. Section~\ref{sec:methodology} presents our data collection process and the empirical design of each research question. Section~\ref{sec:results} provides the results for each research question, followed by Section~\ref{sec:discussion} discussing the results from the previous section. Section~\ref{sec:threats} presents the threats to the validity of the conducted research. Finally, Section~\ref{sec:conclusion} concludes.

\section{Background}
\label{sec:background}
In this section, we start by explaining the concepts of task-based build technologies and artifact-based build technologies. Next, we look into how parallel build and incremental build function in task-based and artifact-based build system technologies. Lastly, we discuss our research within the existing literature on build systems and CI.

\subsection{Task-based Build Technologies}
\begin{figure}
    \centering
    \includegraphics[width=\textwidth]{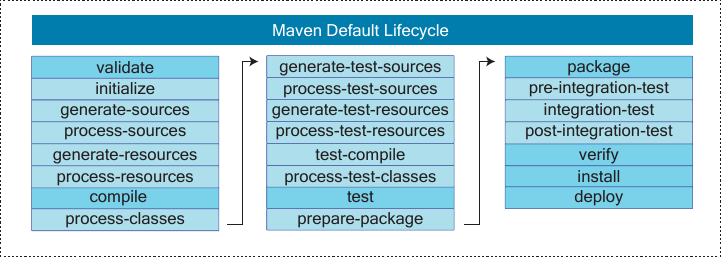}
    \caption{The phases in Maven default lifecycle (The phases with darker backgrounds are the most commonly used ones).}
    \label{fig:maven_default_lifecycle}
\end{figure}

Traditional build systems such as Maven, Ant, and Gradle are task-based build system technologies,
where the build process comprises a set of configurable tasks that depend on each other.
These tasks are specified by developers in build configuration files, instructing the build system on how to carry out the build.
For example, Figure~\ref{fig:maven_default_lifecycle} shows the Maven default build lifecycle,
which handles a project's build and deployment. This lifecycle consists of a sequence of build phases executed in order.
Developers control the Maven project's build process by configuring these phases in the \texttt{pom.xml} file. Figure~\ref{fig:mavenpomfile} shows an example of a Maven \texttt{pom.xml} file, where two plugins are configured. In the example, the \texttt{maven-compiler-plugin} (executed in the \texttt{compile} phase) set the Java version of the source codes and compiled classes to 1.8. The \texttt{maven-jar-plugin} (executed in the \texttt{package} phase) is configured to exclude files under the config directory from the generated jar. 

Task-based build systems are powerful and highly customizable, but they leave too much power to the developers to define and customize these tasks.
The underlying build systems have no idea what these tasks actually do and what they depend on,
which could lead to incorrect build results and poor performance in parallel and incremental builds \citep{fan2020escaping}.
Because the implementation of tasks is obscure to build systems, build systems have to be conservative when scheduling and executing tasks in parallel builds to ensure the correctness of results.
The conservative approach in parallel execution could impact the parallelism of the build process. If not handled carefully, it may even lead to the failure of parallel builds due to potential race conditions \citep{licker2019detecting}.

Furthermore, since build systems might not have complete knowledge of task dependencies, they cannot confidently reuse previous build results as they do not know what the task depends on and if the dependencies have changed.
The unspecified and transitive dependencies in tasks can further contribute to incorrect incremental builds \citep{morgenthaler2012searching} \citep{bezemer2017empirical}.

\definecolor{maroon}{rgb}{0.5,0,0}
\definecolor{darkgreen}{rgb}{0,0.5,0}
\lstdefinelanguage{XML}
{
  basicstyle=\ttfamily,
  morestring=[s]{"}{"},
  morecomment=[s]{?}{?},
  morecomment=[s]{!--}{--},
  commentstyle=\color{darkgreen},
  moredelim=[s][\color{black}]{>}{<},
  moredelim=[s][\color{red}]{\ }{=},
  stringstyle=\color{blue},
  identifierstyle=\color{maroon}
}

\begin{figure}
\begin{lstlisting}[frame=single, language=XML]
<project>
  [...]
  <build>
    [...]
    <plugins>
      <plugin>
        <groupId>org.apache.maven.plugins</groupId>
        <artifactId>maven-compiler-plugin</artifactId>
        <version>3.12.1</version>
        <configuration>
          <source>1.8</source>
          <target>1.8</target>
        </configuration>
      </plugin>
      <plugin>
        <groupId>org.apache.maven.plugins</groupId>
        <artifactId>maven-jar-plugin</artifactId>
        <version>3.3.0</version>
        <configuration>
          <excludes>
            <exclude>**/config/*</exclude>
          </excludes>
        </configuration>
      </plugin>
    </plugins>
    [...]
  </build>
  [...]
</project>
\end{lstlisting}
\caption{An example Maven pom file}
\label{fig:mavenpomfile}
\end{figure}

\subsection{Artifact-based Build Technologies}
In artifact-based build system technologies like Bazel, developers do not explicitly specify the build tasks nor their execution process.
The build system is responsible for configuring, scheduling, and executing tasks.
Artifact-based build systems require developers to declaratively describe in the build file the set of artifacts to be built as well as the dependencies between those and 3rd party components needed to perform a build. This way, developers inform the build system what needs to be built, and the build system figures out how to carry out the building process.
Developers must explicitly specify both external (e.g., third-party libraries, compilers) and internal (e.g., other components in the project) dependencies of the artifacts in the build files.
As the build system knows all the dependencies of the artifacts and has full control over the build process, it can effectively parallelize the tasks and reuse the previous build results while ensuring correctness. Moreover, given that build systems control both the input and tasks of the build process, they also achieve the reproducibility of builds.

Figure~\ref{fig:bazelbuildfile} shows an example of a Bazel build file that contains three Bazel build rules, each describing an artifact. The \texttt{java\_library} build rule in the example compiles the source files located in the same package as the build file, and using an external dependency, \texttt{@maven//:com\_go\-ogle\_guava\_guava}, generates a \texttt{.jar} file. The \texttt{java\_binary} build rule uses \texttt{java-maven-lib}, the name of the aforementioned jar file, as a dependency, and defines an entry point \texttt{com.example.myproject.App} for the jar file to make an executable.

It is worth noting that in Bazel, a test suite is also an artifact, in which the test rule produces an executable with a test runner as the entry point and runs all the tests defined in the test rule. The \texttt{java\_test} build rule in the example specifies the test class, test code, and dependencies (1 internal and 2 external dependencies are imported) of the test code. The executable built by the \texttt{java\_test} build rule loads the test class and executes all test codes. 

\begin{figure}[htp]
\begin{python}[breaklines=true]
load("@rules_java//java:defs.bzl", "java_binary", "java_library")

package(default_visibility = ["//visibility:public"]) 
java_library(
    name = "java-maven-lib",
    srcs = glob(["src/main/java/com/example/myproject/*.java"]),
    deps = ["@maven//:com_google_guava_guava"],
)

java_binary(
    name = "java-maven",
    main_class = "com.example.myproject.App",
    runtime_deps = [":java-maven-lib"],
)

java_test(
    name = "tests",
    srcs = glob(["src/test/java/com/example/myproject/*.java"]),
    test_class = "com.example.myproject.TestApp",
    deps = [
        ":java-maven-lib",
        "@maven//:com_google_guava_guava",
        "@maven//:junit_junit",
    ],
)
\end{python}
\caption{An example of the Bazel build configuration file (https://github.com/bazelbuild/examples)}
\label{fig:bazelbuildfile}
\end{figure}

\subsection{Parallel Build and Incremental Build}
Parallel and incremental builds both are common ways to speed up the build process and are supported by most major build technologies.
Here we discuss the implementation of parallel build and incremental build in Maven and Bazel. 
Since we focus on the duration of the compilation activities during a build, we do not cover the available techniques for test parallelization in these two build technologies. However, since test suites, as everything else in Bazel, are treated as artifacts, the same mechanisms of parallel and incremental build execution apply to test execution within Bazel.

\begin{figure}
\centering
\subfigure(a){\includegraphics[width=0.45\textwidth]{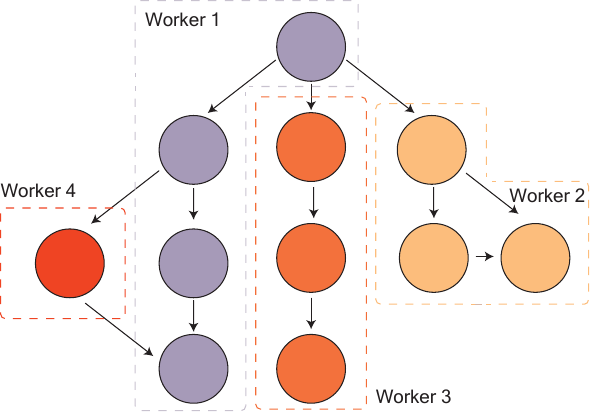}}
\subfigure(b){\includegraphics[width=0.45\textwidth]{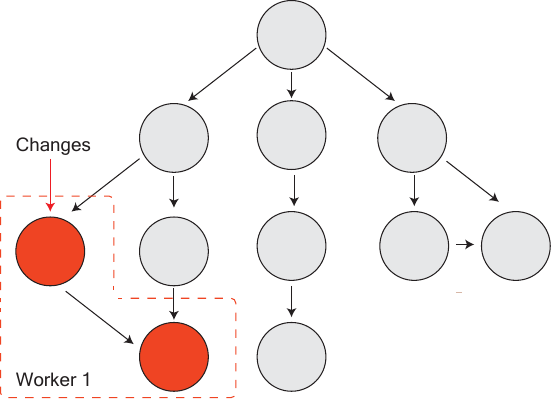}}

\caption{(a) The DAG of the build process with parallel build. (b) The DAG of the build process with incremental build. Each node in the DAG represents a compilation unit in the build.}
\label{fig:parallel_and_incremental_example} 
\end{figure}

When building projects, build systems analyze the external dependencies and internal dependencies of the projects and generate a directed acyclic graph (DAG).
According to this DAG, a build tool determines how to orchestrate the compilation tasks to finish the build process.
Figure~\ref{fig:parallel_and_incremental_example} (a) shows an example of the build process with parallel build, where each node in the DAG represents a compilation unit.
Assuming each node has the same build time \textit{T}, before employing the parallel build, the total build time is 11\textit{T}.
With the parallel build, in the example, build systems can have at most 4 workers running to build the projects, which reduces the build time to 4\textit{T}.

Figure~\ref{fig:parallel_and_incremental_example} (b) illustrates the build process with incremental build. In incremental build, build systems compile only units that have been changed or have direct or transitive dependencies that were changed, whereas reusing the previous build results of other compilation units. In the DAG, the leftmost compilation unit on level three has changed, therefore Bazel builds the compilation unit and every compilation unit depending on it, i.e., two compilation units in the example, with a resulting time of 2\textit{T}.

The major differences between parallel and incremental builds of Maven and Bazel are the accuracy and granularity of the DAG. In Maven, a compilation unit is a module that consists of multiple Java packages \citep{mavendoc}, while a compilation unit in Bazel for Java projects is usually an individual Java package \citep{bazeldoc}. The coarse granularity of Maven tasks in parallel execution reduces its potential for parallelism, however, if the granularity is too small it may increase execution time because of the costs of synchronization of build results between threads \citep{bramas2020improving}. The large granularity of a Maven compilation unit also leads to underutilized dependencies \citep{vakilian2015automated} \citep{jendele2019efficient}. Such underutilized dependencies reduce the effectiveness of incremental builds, as a compilation unit that depends on a very small part of a dependency might need to be rebuilt due to changes in other parts of the dependency, even if the changes are unrelated to the compilation unit.

To generate the DAG, build systems must analyze the external and internal dependencies of the projects. As we discussed before, the task-based build systems do not have enough control over the tasks as they are configured by developers. The uncertainty of the side effects of the tasks results in a less accurate dependency graph and, therefore, more error-prone parallel and incremental builds. As shown in Figure~\ref{fig:bazelbuildfile}, Bazel solves this problem by reserving the control of tasks to itself and requires developers to explicitly write down all dependencies in the build files. However, while it may lead to better performance and correctness in parallel and incremental builds, it also leads to high maintenance costs, since developers must manually maintain the dependency graph in the build files.

\subsection{Related Work}
There are extensive studies investigating the usage of build technologies in open-source projects. \cite{mcintosh2011empirical} studied ten large projects and discovered that maintaining build systems brings 27\% overheads to the development process. \citep{mcintosh2015large} analyzed 177,039 repositories and uncovered that framework-driven build technologies such as Maven are more coupled with source code and require more maintenance effort than low-level build technologies such as Make. \cite{mcintosh2012evolution} and \cite{macho2021nature} studied the evolution and change patterns of build configuration files to understand the co-evolution between build systems and source codes. However, these studies mostly focused on the maintenance activity of the build systems and did not investigate how developers actually use build systems, especially in the context of CI.

Other studies on build technologies and CI mainly focus on CI build breakage and prediction. \cite{rausch2017empirical} performed analysis on CI build failures of 14 open-source Java projects and identified that test failures are the most common errors, while the stability of recent CI builds is the strongest factor influencing the CI build outcome. \cite{zolfagharinia2017not} studied 30 million CPAN builds and discovered the impact of the environment on the CI build outcomes. Several studies have proposed techniques to predict the CI build results and skip CI builds to save build time \citep{xia2017could} \citep{barrak2021builds} \citep{jin2020cost}. Still, none of these studies evaluated how developers use and how much developers exploit the modern performance features of build technologies in CI.

Incremental build also has been getting attention in the literature. The correctness of incremental build relies on an accurate underlying dependency graph.
Several studies investigated the unspecified and redundant dependencies in build systems.
\cite{bezemer2017empirical} studied 4 open-source projects and identified 6 common causes of unspecified dependencies.
\cite{licker2019detecting} proposed a technique to detect the missing dependencies in Make build files.
Some studies discussed the design and principles of an ideal incremental build system.
\cite{erdweg2015sound} designed a build system that provides reliable incremental build with support for dynamic dependencies.
\cite{maudoux2017bringing} discussed why developers do not use incremental builds in CI and proposed criteria for ideal incremental build systems.

Only a few studies examined the usage and benefits of incremental build in open-source projects.
\cite{randrianaina2022benefits} proposed the idea of incremental build of software configurations and conducted a study on 5 open-source projects to evaluate its usefulness. Their results show that an average of 88.5\% of the configurations could be built faster with incremental builds, while 60\% of faster incremental builds are correct.
\cite{ghaleb2021studying} analyzed 513,384 Travis CI builds of 1,270 projects to understand the adaptation of 3rd party dependencies caching in CI. However, their analysis primarily focused on general cache usage in CI rather than caches specific to build results (i.e., incremental build). Moreover, they conducted empirical case studies to investigate the caching adoption in CI. In contrast, we performed experiments to evaluate the performance benefits of incremental build with different caching configurations in CI.

Additionally, the studies on build technologies usually only cover traditional task-based build technologies such as Maven, Ant, and Gradle. However, there is a notable gap in the literature regarding the usage of artifact-based build technologies like Bazel in open-source projects. Apart from the work of Alfadel et al. discussed in the introduction \citep{alfadel2024icse}, several studies from Google have investigated the building and testing process using Bazel in their internal projects (the internal version of Bazel at Google is called "Blaze"). \cite{jendele2019efficient} discussed their work at Google with Bazel and proposed a tool to automatically decompose the compilation units of a Bazel build system to improve the performance of the build process. \cite{wang2021smart} proposed a technique for optimizing Bazel's distributed builds to avoid out-of-memory and deadline-exceeded errors due to the magnitude of Google's huge monolithic codebase. \cite{memon2017taming} also presented their project at Google based on Bazel to reduce the test time by controlling the test workload without compromising quality and by distilling test result data to inform developers. All the above studies focused only on Google's internal projects and did not analyze whether and how developers use Bazel in these projects.

\section{Methodology}
\label{sec:methodology}
In this section, we present the methodology of our empirical study addressing the RQs of the introduction. Specifically, we describe the process of data collection and processing, as well as the experimental setup used to address our research questions. All the scripts and CSV files used in the paper were uploaded to GitHub\footnote{https://github.com/SAILResearch/replication-23-shenyu-bazel\_usage}.

\subsection{Data Collection}

\begin{figure}
  \includegraphics[width=1\textwidth]{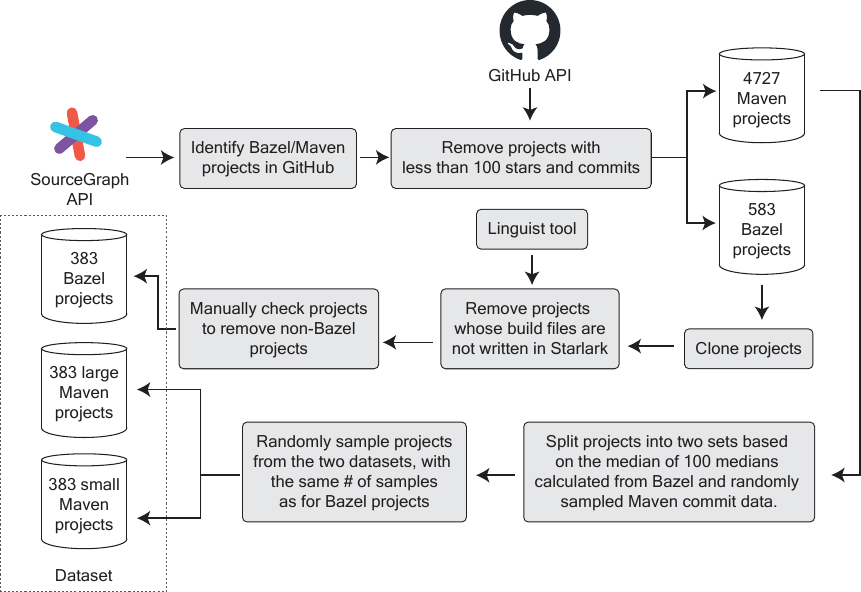}
\caption{The process of data collection}
\label{fig:data_collection}
\end{figure}

Figure~\ref{fig:data_collection} shows an overview of the data collection process. We used Sourcegraph's search API \citep{sourcegraph_2023}, which is a code search engine hosting more than 2 million open-source projects, to collect Bazel projects hosted on GitHub. This is because, as of the day that we collected the data, February 13th, 2023, GitHub’s search API does not provide good support to search for projects containing Bazel build files. Since Maven is the previous state-of-the-art (task-based) build technology studied in research \citep{mcintosh2012evolution}\citep{mcintosh2015large}, we also analyze Maven projects and compare the usage of Maven and Bazel in open-source projects. 

Conventionally, Bazel projects have a build configuration file named \textit{BUILD} or \textit{BUILD.bazel} in their root directory \citep{bazeldoc}, while Maven projects have a build file \textit{pom.xml} in their root directory \citep{mavendoc}. We used the query \texttt{select:repo (file:\^{}BUILD(.bazel)?\$) count:10000} to search for Bazel projects and the query \texttt{select:repo (file:\^{}pom.xml\$) count:10000} to search for Maven projects in Sourcegraph's API. We limited the number of results to 10,000 in the queries to optimize the query time. The returned results are sorted by the number of stars, and, in both results, the last result had less than 100 stars. We then used the GitHub API to remove the remaining projects with less than 100 stars and commits from the dataset of 10,000 projects to eliminate trivial projects, which left us with a total of 583 Bazel projects and 4,727 Maven projects.

Because \textit{BUILD} is a very common file name (for example, other build technologies like Pants and Buck also used \textit{BUILD} as the default name of their build configuration files), we then cloned Bazel projects and employed a tool named Linguist \citep{linguist}, to detect the programming language used in the build file to filter out projects whose \textit{BUILD} or \textit{BUILD.bazel} files are not written in Starlark, which is a Python dialect language used for Bazel build configuration files \citep{starlark}. As the results returned by Linguist may not be entirely reliable, we manually checked the remaining 393 Bazel projects, and we found 10 non-Bazel projects in the dataset. As shown in Figure~\ref{fig:data_collection}, we retrieved 383 Bazel projects after applying the above steps.

To compare the usage of build systems between Bazel and Maven projects, we divided and sampled the Maven projects into two groups based on their number of commits. Due to the disparity in the number of projects within the Bazel (383 projects) and Maven (4,727 projects) groups, to compute a fair threshold for the grouping and the subsequent comparison, we generated 100 random samples of 383 Maven projects, matching the number in the Bazel group. We then calculated the median number of commits for the projects in each sample by aggregating the commit data from both the Bazel group and the randomly selected Maven projects. This resulted in 100 median values. The median of these 100 median values, which is 731.25, was then used as the final threshold for dividing Maven projects, resulting in 2,750 small Maven projects (less than 731.25 commits) and 1,977 large Maven projects (more than 731.25 commits). Subsequently, we randomly sampled 383 projects from the two groups, i.e., the same number of samples for each group as the size of the Bazel group.

\subsection{Research Questions}

\subsubsection{RQ1: To what extent do developers use parallelization and incremental build (cache) features in CI builds?}

\paragraph{Motivation} The performance of build systems greatly influences the successful implementation of CI processes, and long build times could result in less frequent software delivery \citep{maartensson2017continuous}. While Bazel provides rich and powerful features to developers to speed up their build process, it is unclear how developers use these features in their projects, especially in the context of CI. \cite{alfadel2024icse} also suggest that if the perceived benefits of Bazel's feature richness do not outweigh its maintenance costs, it could lead to the abandonment of Bazel.

In RQ1, we investigate the extent to which Bazel's parallelization and incremental build features are employed in CI builds. Before this, we first examine how common Bazel is used in CI in general. Then, we analyze whether developers use parallelization and incremental build features in their CI builds. We compare the results to the results of Maven projects to understand the differences between traditional and modern build systems.

\paragraph{Approach} To understand the parallelization and incremental build (cache) features of Bazel in open-source projects, we leverage the following approaches.

\begin{figure}
  \includegraphics[width=1\textwidth]{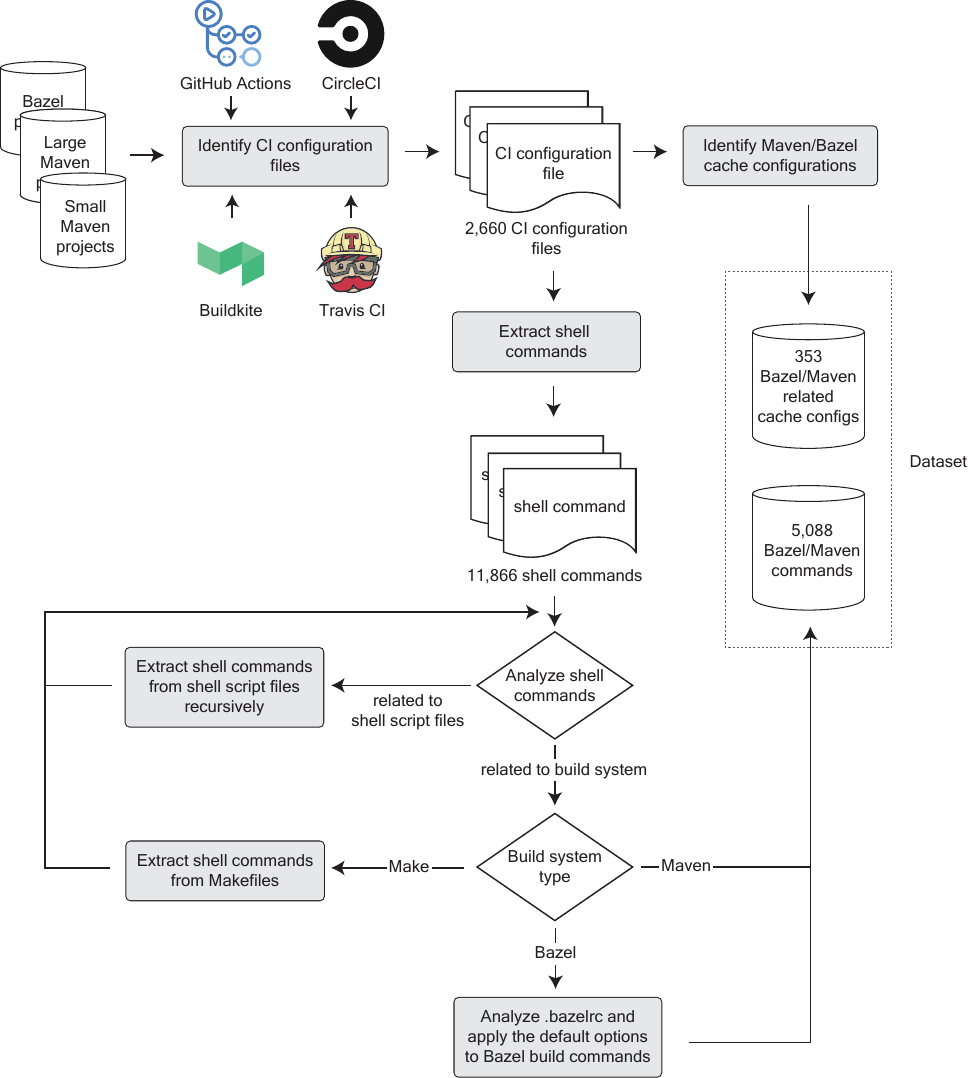}
\caption{The process of analyzing CI configuration files of the projects adopting Bazel or Maven build systems.}
\label{fig:ci_configs_process}  
\end{figure}

\noindent
\newline\textbf{Identify CI Configuration Files}

Figure~\ref{fig:ci_configs_process} shows an overview of the process of analyzing CI configuration files of the projects. \cite{golzadeh2022rise} analyzed 91k GitHub repositories of active npm packages, which revealed that GitHub Actions, Travis CI, and Circle CI are the three most widely used CI services among these repositories. Therefore, for our analysis, we focused on these three CI services, along with Buildkite, as Buildkite is adopted by many Bazel projects. Projects not using these four CI services were excluded from the analysis of CI usage. Since we cannot know the entire population of CI-using Bazel projects in GitHub directly through Sourcegraph's API, we first sampled from the entire population of Bazel projects, then obtained CI-using Bazel projects from this sample (see below). To ensure fairness, we employed the same approach for sampling Maven projects. 

We went through these services' documentation to determine the file paths and specifications of the CI configuration file. Then, to check if a CI service is used by a project, we cloned all projects in the datasets and used the regular expressions listed in Table~\ref{table:config_paths} to search for CI configuration files under the root directory of projects. After examining the 1,149 open-source projects in our three datasets, we identified 289 Bazel, 284 large Maven, and 212 small Maven projects that used the four CI services and 2,660 CI configuration files in these projects. Notably, the number of CI configuration files is much higher than the number of identified projects. This is because many projects use multiple CI services and because GitHub Actions projects usually create multiple configuration files.

\begin{table}
\centering
\caption{Regular Expressions for Matching CI Configuration File Paths}
\label{table:config_paths}
\begin{tabularx}{\textwidth}{l|X}
\toprule
CI Service & Regular Expression \\
\midrule
GitHub & 
\verb@\.github/(workflows|actions)/.*\.ya?ml$@ \\
CircleCI &
\verb@\.circleci/.*\.ya?ml$@ \\
Travis CI &
\verb@\.travis.ya?ml$@  \\
Buildkite &
\verb@\.buildkite/.*\.ya?ml$@ \\
Buildkite (With DSL)* &
\verb@\.bazelci/.*\.ya?ml$@ \\
\bottomrule
\end{tabularx}
\hspace{1cm}
\begin{tabularx}{\linewidth}{XX}
    \multicolumn{2}{X}{* Projects maintained by Bazel developers use a DSL to set up their CI workflows on Buildkite: https://github.com/bazelbuild/continuous-integration} \\   
\end{tabularx}
\end{table}

\noindent
\newline\textbf{Extract Build Commands}

Since build systems like Bazel are not always invoked directly by CI, but can be used within shell scripts, we also needed to analyze the build system-related commands directly or indirectly used in CI configuration files. For this, we followed the specifications of the CI configuration files to extract the shell commands executed in the workflow. The locations of shell commands in CI configuration files are shown in Table~\ref{table:shell_command_paths}. We extracted a total of 11,866 shell commands from the CI configuration files. Additionally, instead of using shell commands to run the build system in CI workflow directly, some projects execute shell script files in their workflows and run the build system within the shell script files. So, for each project, we collected all the files whose name ends with \texttt{.sh} and analyzed those whose names appeared in the shell commands, identifying 1,157 shell script files used in the CI configuration files.

\begin{table}
\centering
\caption{Regular Expressions for Matching Build Commands}
\label{table:build_commands_match}
\begin{tabularx}{\textwidth}{l|X}
\toprule
Build Tool & Regular Expression \\
\midrule
Bazel & \begin{itemize}[label={}, nosep, leftmargin=*, before=\vspace{-0.6\baselineskip}, after =\vspace{-\baselineskip}]
    \item \verb@.*bazel[\"w]? (.+)@
    \item \verb@.*bazelisk (.+)@
\end{itemize} \\
Maven &
\begin{itemize}[label={}, nosep, leftmargin=*, before=\vspace{-0.6\baselineskip}, after =\vspace{-\baselineskip}]
    \item \verb@.*mvnw? (.*)@
\end{itemize} \\
Make &
\begin{itemize}[label={}, nosep, leftmargin=*, before=\vspace{-0.6\baselineskip}, after =\vspace{-\baselineskip}]
    \item \verb@.*make (.*)@
\end{itemize} \\
\bottomrule
\end{tabularx}
\end{table}

Next, we used regular expressions shown in Table~\ref{table:build_commands_match} to identify build system-related commands from the shell commands and shell script files. Initially, we only examined build commands for Maven and Bazel. However, we discovered that some projects use Make to execute other build systems. Hence, we used another regular expression to match Make build commands and analyzed the shell commands run by the Make targets, including their prerequisites, in the build commands. Furthermore, we analyzed the .bazelrc file in the root directory of Bazel projects, which contains the default command-line options of Bazel commands. Finally, we extracted a total of 5,088 build commands (shell commands that run build systems) associated with Bazel or Maven.

\begin{table}
\centering
\caption{Locations of Shell Commands in CI Configuration Files}
\label{table:shell_command_paths}
\begin{tabularx}{\textwidth}{lX}
\toprule
CI Service & Location \\
\midrule
GitHub & 
\begin{itemize}[nosep, leftmargin=*, before=\vspace{-0.6\baselineskip}, after =\vspace{-\baselineskip}]
    \item \verb@jobs.<job_id>.steps[*].run@
    \item \verb@runs.steps[*].run@
\end{itemize} \\
\cmidrule{1-2}
CircleCI &
\begin{itemize}[nosep, leftmargin=*, before=\vspace{-0.6\baselineskip}, after =\vspace{-\baselineskip}]
    \item \verb@jobs.<job_name>.steps[*].run@
    \item \verb@commands.<command>.steps[*].run@
\end{itemize} \\
\cmidrule{1-2}
Travis CI &
\begin{itemize}[nosep, leftmargin=*, before=\vspace{-0.6\baselineskip}, after =\vspace{-\baselineskip}]
    \item \verb@script@
    \item \verb@jobs.include[*].script@
    \item \verb@jobs.exclude[*].script@
    \item \verb@install@
    \item \verb@deploy@
\end{itemize} \\
\cmidrule{1-2}
Buildkite &
\begin{itemize}[nosep, leftmargin=*, before=\vspace{-0.6\baselineskip}, after =\vspace{-\baselineskip}]
    \item \verb@steps[*].command@
    \item \verb@steps[*].commands@
\end{itemize} \\
\cmidrule{1-2}
Buildkite (With DSL) &
\begin{itemize}[nosep, leftmargin=*, before=\vspace{-0.6\baselineskip}, after =\vspace{-\baselineskip}]
    \item \verb@platforms.<id>@
    \item \verb@tasks.<id>@
\end{itemize} \\
\bottomrule
\end{tabularx}
\end{table}

\noindent
\newline\textbf{Analyze Build Commands}

To assess the usage of parallelization, we inspected the build command options. By default, Bazel has parallelization enabled, and therefore all Bazel projects are presumed to use parallelization unless they use the \texttt{-j} or \texttt{--job} option to specify the number of concurrent jobs to run. On the other hand, parallelization is not enabled by default in Maven, and thus all Maven projects are presumed not to use the parallelization feature unless they use the \texttt{-T} or \texttt{--threads} option and specify a value greater than 1. We analyzed the values of these command options to obtain the desired parallelism of each build system.

Incremental build is an effective technique for reducing build time and is often achieved by sharing caches in CI. Although both build systems and CI services support cache sharing between CI builds, their sharing mechanisms differ in certain aspects. The cache provided by CI (General-Purpose CI Cache) needs to be build system technology-independent, hence typically leverages general-purpose file or object storage. In this case, the build system reads from and writes to a local file path that essentially corresponds to a file stored in the General-Purpose CI Cache. Conversely, the build system remote cache (Build-Tool-Specific Cache) is specifically designed for a given build tool and, therefore only caches files deemed necessary for the build by the build tool using the tool's official API. In addition, the Build-Tool-Specific Cache usually incorporates specialized optimizations (e.g., cache integrity validation, gRPC support) to improve performance and security, whereas the General-Purpose CI Cache lacks such tailored optimizations.

Therefore, for the usage of incremental build in CI, we examined both Build-Tool-Specific Cache and General-Purpose CI Cache. We first analyzed CI configuration files to check the usage of the General-Purpose CI Cache. We referred to the documentation provided by the CI services to locate the configuration related to cache usage. Table~\ref{table:local_cache_paths} illustrates the areas we examined to identify the General-Purpose CI Cache usage. Once we identified the cache-related configurations from the files, we examined the paths of the cached files in the cache configurations. We evaluated whether these paths contain the substring \texttt{.cache/bazel} or \texttt{.m2} to determine if they facilitate caching for Bazel or Maven.

Then, we examined the Build-Tool-Specific Cache usage. We looked at the build command options and build configuration files. For Bazel projects, we looked for the \texttt{-{}-remote\_cache} option in build commands to determine if they are using the Build-Tool-Specific Cache. For Maven projects, cache sharing between different machines is supported by the Maven Build Cache Extension \citep{mavendoc}, so we inspected the \texttt{pom.xml} in the projects to check if they are using this extension and determined the Build-Tool-Specific Cache usage of Maven projects.

\begin{table}
\centering
\caption{Locations of Cache-related configuration in CI Configuration Files}
\label{table:local_cache_paths}
\begin{tabularx}{\textwidth}{lX}
\toprule
CI Service & Location \\
\midrule
GitHub & 
\begin{itemize}[nosep, leftmargin=*, before=\vspace{-0.6\baselineskip}, after =\vspace{-\baselineskip}]
    \item \verb|jobs.<job_id>.steps[uses=actions/cache]|
    \item \verb|jobs.<job_id>.steps[uses=actions/setup-java]|
\end{itemize} \\
\cmidrule{1-2}
CircleCI &
\begin{itemize}[nosep, leftmargin=*, before=\vspace{-0.6\baselineskip}, after =\vspace{-\baselineskip}]
    \item \verb@jobs.<job_name>.steps.restore_cache@
    \item \verb@jobs.<job_name>.steps.save_cache@
\end{itemize} \\
\cmidrule{1-2}
Travis CI &
\begin{itemize}[nosep, leftmargin=*, before=\vspace{-0.6\baselineskip}, after =\vspace{-\baselineskip}]
    \item \verb@cache@
\end{itemize} \\
\cmidrule{1-2}
Buildkite &
\begin{itemize}[nosep, leftmargin=*, before=\vspace{-0.6\baselineskip}, after =\vspace{-\baselineskip}]
    \item \verb@steps[*].plugins[gencer/cache]@
\end{itemize} \\
\cmidrule{1-2}
Buildkite (With DSL) &
N/A
\\
\bottomrule
\end{tabularx}
\end{table}

\subsubsection{RQ2: What is the impact of Bazel parallelization on the build performance?}

\paragraph{Motivation} Bazel claims to provide better parallelism than traditional task-based build technologies, and parallel build is by default enabled in Bazel. However, developers may not be able to fully utilize this capability provided by Bazel, for a variety of reasons, in CI. The default number of cores of the free tiers of the three most popular CI services in GitHub is only 2 \citep{githubactionsdoc} \citep{circlecidoc} \citep{traviscidoc}, so even if Bazel provides powerful parallelism, there is an upper limitation to the parallelism Bazel may achieve in most CI environments. Furthermore, Bazel uses the dependency graph of the projects to determine how to build projects in parallel. If the project's structure is not well-organized, the parallelization might be underutilized. 

In this RQ, we build projects with different parallelism degrees to understand the extent to which developers exploit the parallelization feature of Bazel in open-source projects and the optimal parallelism achievable for these projects.

\begin{figure}
  \includegraphics[width=1\textwidth]{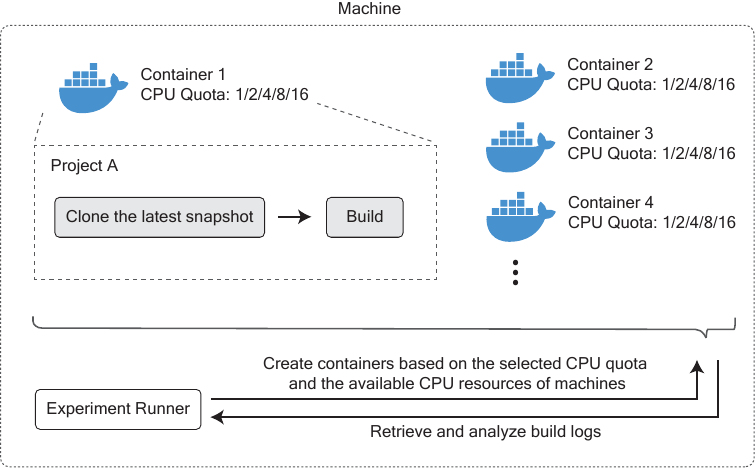}
\caption{Experiments for parallelization usage of Bazel (RQ2).}
\label{fig:parallelization_experiment_setup}  
\end{figure}

\paragraph{Approach} To evaluate the utilization of the parallelization of Bazel, we performed builds with five different parallelism degrees on the latest snapshots of 383 Bazel projects obtained from GitHub in the data collection section. Then, we excluded projects with failed builds during the experiments and analyzed the build logs of the remaining 70 buildable projects to measure how much they can utilize the parallelism. 

\noindent
\newline\textbf{Experiment Setup}

Figure~\ref{fig:parallelization_experiment_setup} illustrates an overview of our experiments. The experiments conducted in RQ2 were executed on 6 Google Cloud Platform e2-standard-16 instances, each equipped with 16 vCPUs, 64 GB of memory, and a 100 GB SSD disk.

We took the following actions to mitigate any potential interference from external factors. The machines employed for the experiments were exclusively dedicated to the study. We ran each experiment within a separate container that was specifically created for this experiment and was removed after the experiment ended. To control the CPU resources allocated to containers, we employed the \texttt{-{}-cpu} option when starting docker containers, which limited the number of cores available for containers. Additionally, we controlled the number of running containers based on the available CPU resources on the servers and the CPU quota required by the containers to ensure the optimal utilization of CPU resources for each container. Each project was built with five different parallelism degrees: 1, 2, 4, 8, and 16. To control the effect of noise between different builds, for each parallelism degree, we performed the build 10 times and used the median build duration value for analysis.

We analyzed the Bazel build commands identified previously in RQ1 and found the percentages of projects using Bazel for compilation (via the \texttt{build} subcommand) and for testing (via the \texttt{test} subcommand) in CI to be similar, i.e., 81.52\% and 79.15\%, respectively. In addition to the fact that compilation and test execution in Bazel build systems leverage the same mechanisms as well (as explained in Section~\ref{sec:background}), we decided to specifically focus on compilation targets in our experiments, leaving test execution for future work.

In Bazel, the targets to be built are specified by developers and can differ across projects. To determine the build targets for our experiments, we first manually examined the Bazel-related build commands extracted from CI configuration files in RQ1. In cases where some projects did not use Bazel in CI or used different Bazel commands instead of the \textit{build} subcommand in CI, we manually looked at these projects' documentation to identify the build targets. If no target could be identified through this step, we inspected the build files at the root of the projects to identify any top-level buildable targets defined in those files. In such cases,  we used the \texttt{//:all} target in the experiments, which builds all the targets defined at the top level. If no top-level build target was defined, we used the \texttt{//...} target to build all targets in the projects. As a result, we used the specific targets for 25 projects, the \texttt{//:all} targets for 249 projects, and the \texttt{//...} targets for 109 projects. Additionally, considering the difference in Bazel versions across projects, in the experiments, we used Bazelisk\footnote{https://github.com/bazelbuild/bazelisk}, a commonly used tool in the Bazel community, to automatically identify and download the suitable Bazel version for each project.

For each experiment, we cloned the project's latest snapshot into the container and performed the build. To measure the build time for each project, we used a script to analyze the build logs obtained from Bazel, which show the duration of the whole build process. In total, we conducted 3,500 experiments on the 70 Bazel projects, out of the 383 Bazel projects that could be built without errors. The remaining projects that did not build successfully were excluded from the analysis.

Parallelism should especially benefit long-duration builds. However, it is not entirely clear if this is indeed the case. Therefore, we sorted the 70 buildable projects based on their median baseline build duration (the median build time of builds at parallelism 1) and then divided them into three groups: short- (23 projects), medium- (23 projects), and long-build duration projects (24 projects). Afterward, we compared the results between groups to understand how to choose parallelism when building projects. 

\begin{figure}
  \centering
  \includegraphics[width=0.8\textwidth]{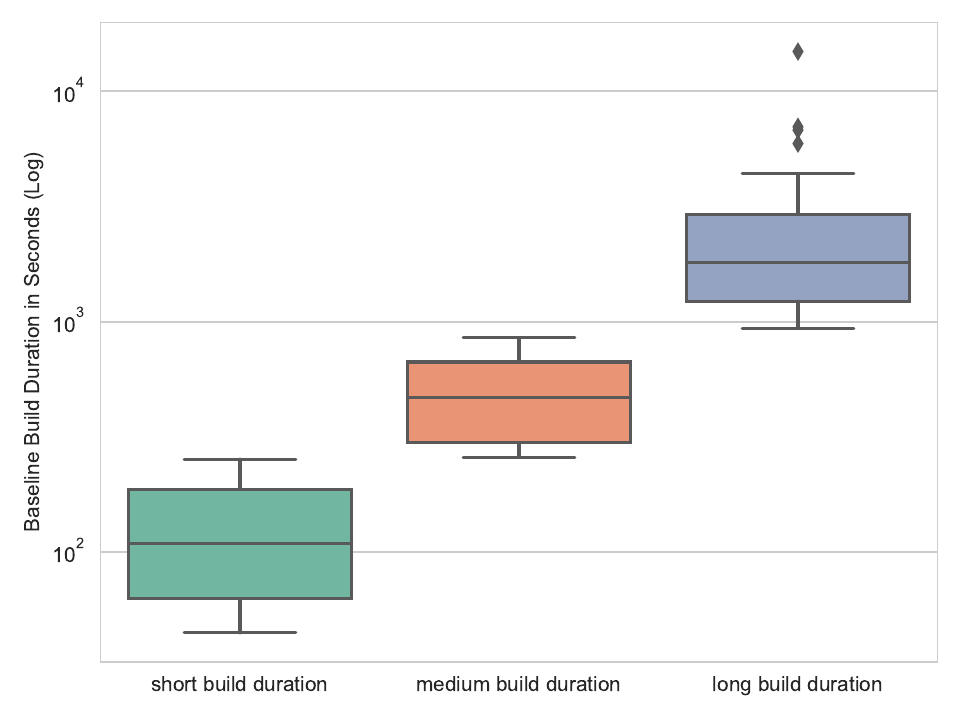}
\caption{Distribution of the median baseline build times of the parallelization experiments.}
\label{fig:parallelization_baseline_build_time}  
\end{figure}

Figure~\ref{fig:parallelization_baseline_build_time} shows the distribution of the median baseline build times for each group. The minimum, maximum, and median baseline build times for short-build duration projects are 44.74s, 253.41s, and 108.88s, respectively. The corresponding figures for medium-build duration projects are 258.21s, 850.49s, and 467.78s, respectively, and for long-build duration projects are 935.84s, 14836.71s, and 1801.53s, respectively.

\noindent
\newline\textbf{Dependency Graph Analysis}

Considering that the structure of projects might greatly influence the parallelization efficiency, we also examined both the structure and the granularity of the dependency graph to understand their impact on parallelization efficiency. 

Bazel analyzes the artifacts and their dependencies defined in the build files to generate a dependency graph used to determine how to conduct the build process. In the dependency graph, the nodes are artifacts and the edges represent the dependencies between artifacts. To extract the dependency graph, we employed a script to run the Bazel \textit{cquery} command for each project. As the generated dependency graph lacks input information for artifacts, we also executed the Bazel \textit{aquery} command to query the paths of input files being compiled for each artifact.

We aimed to understand how the structure of projects affects parallelization efficiency. Therefore, in the network analysis of the Bazel dependency graph, we specifically focused on the measures related to coupling and cohesion. 

The coupling of software systems can be summarized by the degree distribution of their network \citep{savic2019complex}. The degree distribution of highly coupled software networks usually follows the power-law distribution, where most nodes only have a few connections to other nodes and a few nodes have many connections \citep{taube2011can}. However, while fitting power-law distribution to the degree distribution is a common way to evaluate the coupling of software systems, \cite{broido2019scale} found that power-law degree distribution is rare in real-world networks, and in 88\% of analyzed networks, a log-normal fits the degree distribution as well as or better than a power law. \cite{savic2019complex} also found that high coupling in real software systems cannot be accurately modeled by power-law distributions. Therefore, we picked three coupling-related measures used in \cite{savic2019complex}'s study (shown in Table~\ref{table:network_measures}) to assess the coupling of the dependency graph. 

To assess the cohesion of the dependency graph, we investigated whether the network displays the small-world property, characterized by a high clustering coefficient and a low average shortest path. The clustering coefficient gauges the extent to which nodes in a network tend to be neighbors, while the average shortest path computes the average number of steps along the shortest path between every pair of nodes in a network. Using both the average shortest path length and average clustering coefficient, we can identify the cohesiveness of software systems \citep{chong2015analyzing}. Since large networks tend to have smaller cluster coefficients and bigger average shortest path lengths, these two measurements are normalized by the corresponding values of the randomly generated Erdos-Renyi network that has the same number of nodes and degrees. In addition to these two metrics, we also analyze the size of the largest weakly connected component in the network. A weakly connected component refers to a subset of the original graph in which all nodes are interconnected through some path, ignoring the direction of edges. The size of the latest weakly connected component reflects the overall cohesiveness of a software project \citep{savic2019complex}.

Next, we examined the granularity of nodes (artifacts) within the dependency graph. Using the paths of input files collected in earlier steps for each artifact, we employed an additional script to calculate the number of lines in each input file. We used the total number of lines of input files as the size of each node. Then, we calculated and used the average node size to measure the overall granularity of the dependency graph. Table~\ref{table:network_measures} shows the complete list of metrics.

\begin{table}
\centering
\caption{Network Measures of the Dependency Graph}
\label{table:network_measures}
\begin{tabularx}{\textwidth}{lX}
\toprule
Measures & Description \\
\midrule
\multicolumn{2}{l}{\textbf{Coupling Measures}} \\
Mean Total Degree \\ (MTD) & The average total degree (i.e., the number of edges connected to a node) of nodes. A high total degree suggests a highly coupled software system. \\
In-degree Skewness \\ (INS) & The skewness of the nodes' in-degree (i.e., the number of edges coming into a node) distribution. A high and positive skewness in in-degrees indicates that a few nodes possess notably high in-degrees, while others exhibit low in-degrees, implying potential coupling. \\
Out-degree Skewness \\ (OUTS) & The skewness of the nodes' out-degree (i.e., the number of edges coming out of a node) distribution. Similar to in-degree skewness, a high and positive skewness in out-degrees indicates that a few nodes possess notably high out-degrees, while others exhibit low out-degrees, which also implies potential coupling. \\
\multicolumn{2}{l}{\textbf{Cohesion Measures}} \\
Percentage of Nodes in the Largest \\ Weakly Connected Component (PWCC) & The percentage of nodes in the largest weakly connected component. It measures the overall cohesiveness. \\
Average Clustering Coefficient \\ (CC) & It measures the extent to which nodes in a network tend to be neighbors. \\
Average Shortest Path Length \\ (ASPL) & The average number of steps along the shortest path between every pair of nodes in a network. \\

\multicolumn{2}{l}{\textbf{Granularity Measures}} \\
Number of Nodes \\ (NNODE) & The number of nodes in the dependency graph. \\

Mean Node Size \\ (MNSIZE) & The average node size of the dependency graph. The node size is defined as the sum of lines across all input files associated with the node (artifact). \\
\bottomrule
\end{tabularx}
\end{table}

Our approach to analyzing the dependency graph measures is similar to \cite{bettenburg2010studying}'s study, i.e., we used a linear regression model to investigate the statistical connections between these metrics (independent variables) and the speedup at various levels of parallelism (dependent variable). To mitigate the skewing effects, we applied log transformation to the independent variables that show skewness. We first conducted a correlation analysis to examine the potential inter-correlation between the independent variables. We removed the independent variable with the highest variance inflation factors \citep{weisberg2005applied} from the model and recomputed the variance factors until none had an inflation factor exceeding 5. During the multi-collinearity analysis, the NNODE independent variable is removed from the model. Due to space constraints, the detailed results table is available online \footnote{https://github.com/SAILResearch/replication-23-shenyu-bazel\_usage}. Following that, we constructed multiple linear regression models to evaluate the relative impact of each of the three groups of measures on the speedup at different degrees of parallelism.

\subsubsection{RQ3: What is the impact of Bazel incremental build (cache) functionality on the build performance?}

\begin{figure}
  \includegraphics[width=1\textwidth]{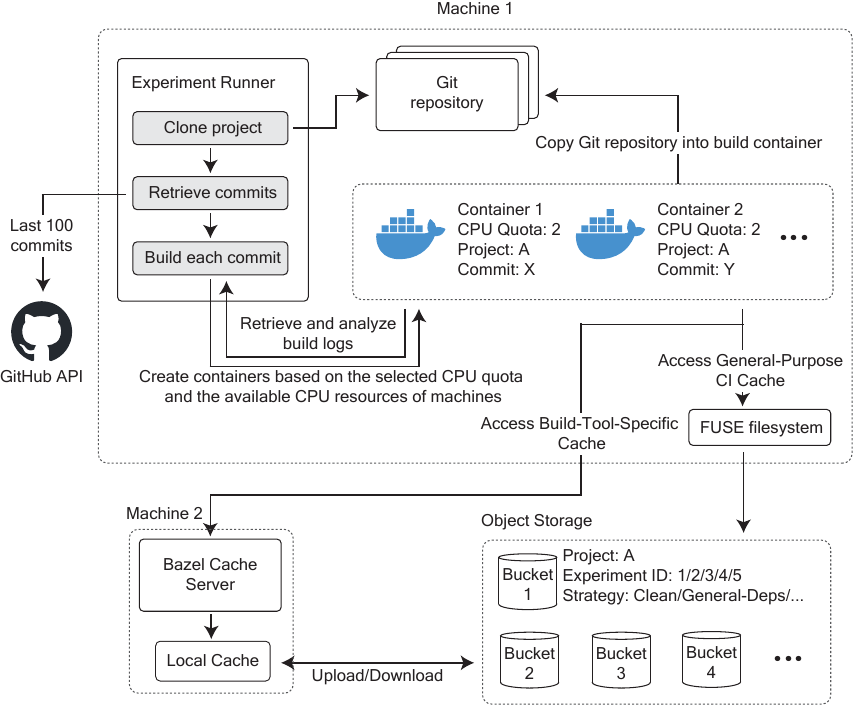}
\caption{Experiments for incremental build utilization of Bazel (RQ3)}
\label{fig:cache_experiment_setup}  
\end{figure}

\paragraph{Motivation} Incremental builds have the potential to significantly reduce build times; however, their utilization in CI services is not widespread \citep{maudoux2017bringing} \citep{randrianaina2022benefits}. Although the results of RQ1 show that Bazel provides strong support for incremental build in CI services, anecdotal evidence\footnote{\url{https://github.com/bazelbuild/bazel/issues/7664}}\footnote{\url{https://github.com/bazelbuild/rules_go/issues/2188}} also shows that Bazel's incremental build may not always improve build performance, especially in CI services, where the process of downloading or uploading caches could be more time-consuming than the actual build itself. 

Thus, in RQ3, we investigate the performance of incremental builds of Bazel in open-source projects in the context of CI and explore the best practices for developers to leverage the incremental build features of Bazel.

\paragraph{Approach} Bazel's incremental build in CI services is achieved by sharing caches between CI builds. There are 3 types of caches in Bazel: Analysis Cache (the cache of dependency graphs), Repository Cache (the cache of external dependencies), and Action Cache (the cache of build results). For our study, we focused on the latter two types of cache since the Analysis Cache is usually stored in memory and not shared in CI builds. Although the Repository Cache alone cannot achieve incremental build, we included it in our study as it is commonly used by build systems in CI services to speed up builds. For clarity, in our study, we refer to the Repository Cache as the \textbf{Dependency Cache} and the Action Cache as the \textbf{Build Result Cache}.

In RQ1, we observed that developers use both General-Purpose CI Cache and Build-Tool-Specific Cache to share files between CI builds. Therefore, we evaluated Bazel's incremental build performance using four cache strategies: No Cache (\textit{Clean}), CI Dependency Cache (\textit{General-Deps}), CI Dependency and Build Result Cache (\textit{General-Deps-and-Results}), and Build-Tool-Specific Dependency and Build Result Cache (\textit{Specific-Deps-and-Results}). To enable each type of cache, we used and combined the following Bazel command-line options: \texttt{-{}-repository\_cache} for CI Dependency Cache, \texttt{-{}-disk\_cache} for CI Build Result Cache, \texttt{-{}-experimental\_remote\_downloader} for Build-Tool-Specific Dependency Cache, and \texttt{-{}-remote\_cache} for Build-Tool-Specific Build Result Cache.

\noindent
\newline\textbf{Experiment Setup}

Figure~\ref{fig:cache_experiment_setup} shows an overview of the incremental build experiments. The incremental build experiments ran on 5 Google Cloud Platform e2-standard-16 instances, and the cache server ran on a separate e2-standard-16 instance equipped with 16 vCPUs, 64 GB of memory, and a 100 GB SSD disk. The object storage used in this RQ is Google Cloud storage. In practice, cache servers are often placed near the build machines to reduce the latency. Therefore, in our experiments, we put the cache and build machines in the same Google Cloud region (e.g., us-central1) to make them geographically close. The cache machine and Google Cloud object storage bucket are also deployed in the same region to make the comparison between \textit{General-Deps-and-Results} and \textit{Specific-Deps-and-Results} strategies fair.

In real-world scenarios, the General-Purpose CI caches are typically stored in object storage, with CI services offering a file system-style API for build systems to access these caches. The most common technique for mounting such remote object storage onto developers' machines and providing access to the storage through file system APIs is FUSE (Filesystem in Userspace). To make our experiments more realistic, we adopted a similar approach and hence employed FUSE to mount the object storage buckets onto the CI machines. During the experiments, Bazel accessed these buckets using the file system APIs.

For the Build-Tool-Specific Cache, we followed the guidance from Bazel's official website for its implementation. To set up the Bazel cache server, we used bazel-remote \citep{bazelremote}. The cache server ran on a separate machine and was only accessible to experiments over the network. While the cache server also used the object storage as its backend storage, it maintained a copy of caches on disk and synchronized the caches with the backend object storage.

In this RQ, we again evaluate the 70 buildable projects identified in RQ2. Similar to RQ2, we used Docker containers to isolate the environments and control the available resources for experiments. Since we focused on incremental build in the CI setting and the default core number is 2 for the free tiers of the three most popular CI services in GitHub, the CPU quota for all experiment containers was set to 2. To avoid the influence of changes to GitHub repositories and speed up the experiments, projects were cloned in advance onto the machines before the start of their experiments. 

In the experiments, we obtained the most recent 100 commits within the past 1,000 days for each project. We sequentially built these commits in their chronological order using the 4 cache strategies. For each project, when conducting experiments with a specific cache strategy, we built commits from the oldest to the newest, excluding the \textit{Clean} strategy. Due to the significantly longer time required for the \textit{Clean} strategy to complete, we built projects using this strategy every 5 commits instead. To make the experiments more realistic, each commit was individually built in a dedicated container. The pre-cloned projects were copied into the containers and checked out to the respective commit for building. We repeated the experiments for each strategy 5 times and used the median value for analysis. We used different cache locations during each repetition of the experiments for each strategy to avoid the impact of caches from other repetitions. As the Bazel version might differ between commits in the project, similar to RQ2, we also used Bazelisk in the experiment to download the suitable version of Bazel for each commit. We employed the build targets identified in RQ2 to build the projects. After the experiments were completed, a script was used to extract the build duration. Given that when the Build Result Cache is enabled in the Bazel build, the build logs show the total number of processes executed in the build process and the number of processes that hit caches, the script we employed also extracted these two values to calculate the cache hit rates.

\begin{figure}
  \centering
  \includegraphics[width=0.8\textwidth]{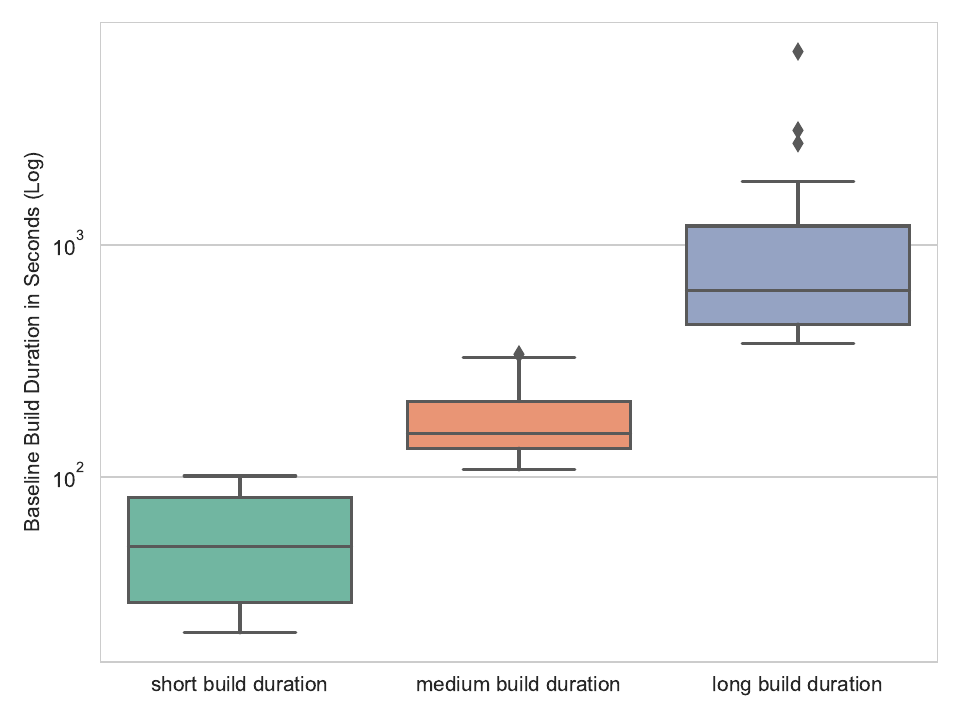}
\caption{Distribution of the median historical baseline build times of the incremental build experiments.}
\label{fig:cache_baseline_build_time}  
\end{figure}

In total, we ran 102,232 experiments on the 70 buildable projects. In contrast to RQ2, here we used the build duration of the \textit{Clean} builds to categorize the projects, as we used the builds using the \textit{Clean} strategy as the baseline. Moreover, because the build duration of a single commit cannot reflect the build duration throughout the history of the projects, we used the median build duration of all \textit{Clean} strategy builds to sort and divide projects into three groups, which are short- (23 projects), medium- (23 projects), and long-build (24 projects) duration projects. Figure~\ref{fig:cache_baseline_build_time} illustrates the median baseline build time distribution of short- (min: 21.25s, max: 100.94s, median: 50.07s), medium- (min: 107.25s, max: 337.82s, median: 155.81s), and long-build (min: 377.28s, max: 6827.24s, median 634.39s) duration projects.

\section{Results}
\label{sec:results}
\newcolumntype{Y}{>{\centering\arraybackslash}X}

In the following, we present the findings of our empirical study for each research question.

\subsection{RQ1: To what extent do developers use parallelization and incremental build (cache) features in CI builds?}

\begin{figure}
  \includegraphics[width=1\textwidth]{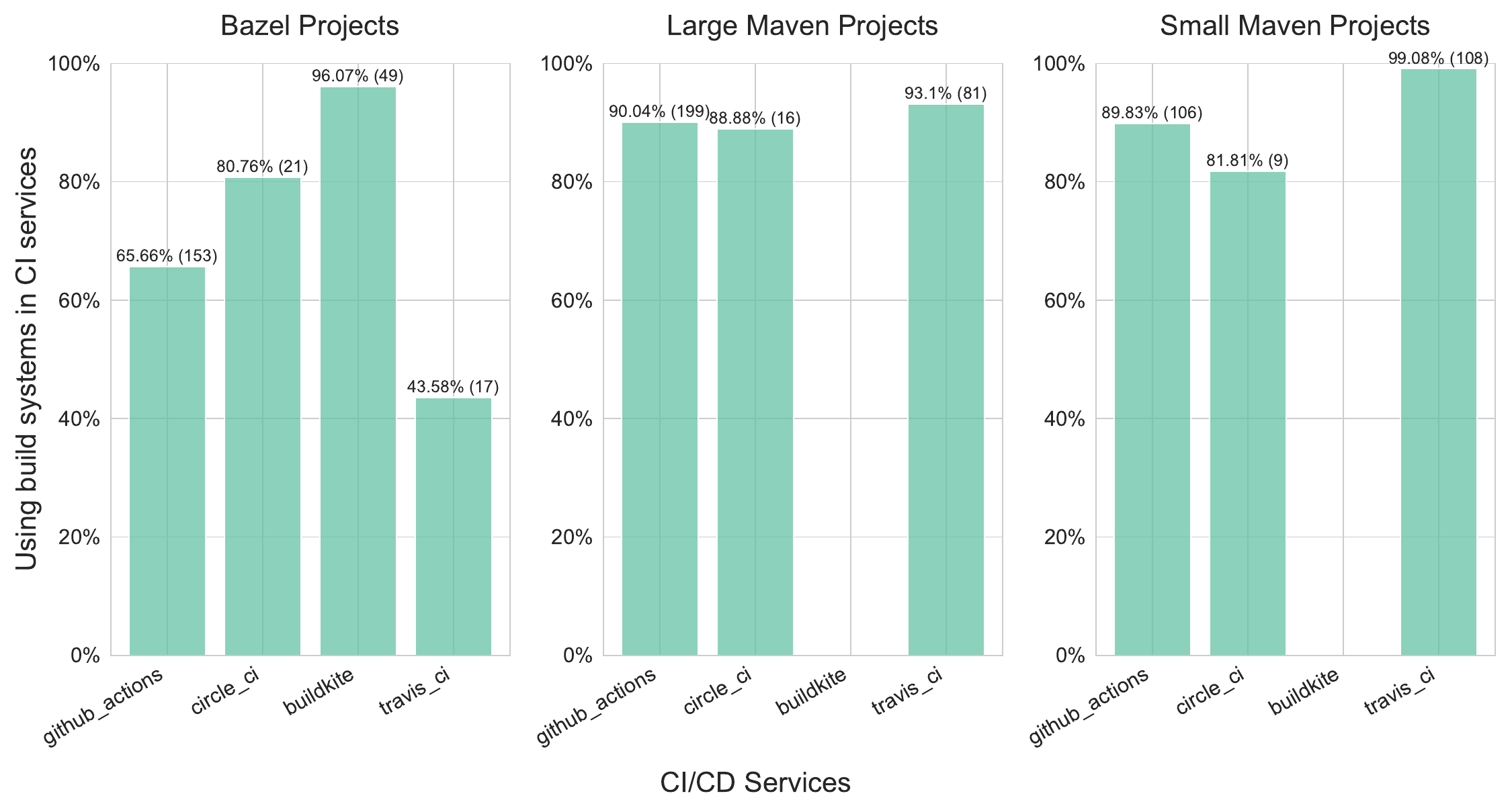}
\caption{The percentage of projects in the dataset adopting a CI service and using the Maven or Bazel build system in CI. A project may be counted multiple times if it uses multiple CI services.}
\label{fig:ci_tool_usage}
\end{figure}

\textbf{Approximately 31.23\% of Bazel projects do not use Bazel in their CI services. In contrast, the corresponding figures for large Maven projects and small Maven projects are 9.48\% and 6.30\%, respectively.} Figure~\ref{fig:ci_tool_usage} shows the usage of build systems in the four studied CI services. Bazel projects exhibit the lowest CI build system usage in Travis CI (43.58\%), GitHub Actions (64.80\%), and Circle CI (80.76\%), except for Buildkite, since the latter is Bazel-specific. Conversely, large Maven projects demonstrate the highest CI build system usage for GitHub Actions (90.04\%) and Circle CI (88.88\%), while small Maven projects
demonstrate the highest CI build system usage for Travis CI (99.08\%).

By manually examining the CI configuration files of the Bazel projects that show no usage of Bazel in the four studied CI services, we found that 53.13\% of them adopted multiple build systems and used other build systems to build in CI services. For example, \textit{sogou/srpc} uses CMake to build the project in CI, while \textit{go-resty/resty} uses Go's toolchains. We looked at these projects further to understand the rationale behind maintaining multiple build systems, as it brings more maintenance overhead \citep{suvorov2012empirical}. We discovered that for projects that use multiple build systems, around half of them (47.46\%) are C++ projects that use CMake first and adopt Bazel later. 

We examined the pull requests that introduced Bazel in these C++ projects and found that the compatibility issues between non-Bazel C++ projects and Bazel C++ projects are the reason why developers adopted Bazel in the projects. A non-Bazel C++ project cannot be easily used as a library of Bazel C++ projects. Developers often have to inject Bazel build configuration files that they write themselves into the non-Bazel C++ projects so that these projects can be used as libraries of Bazel C++ projects. Therefore, projects such as \textit{oneapi-src/oneTBB}\footnote{https://github.com/oneapi-src/oneTBB/pull/442}, \textit{AcademySoftwareFoundation/openexr}\footnote{https://github.com/PCRE2Project/pcre2/pull/136}, despite using CMake as their primary build system, introduced Bazel so these projects can be easily integrated into other Bazel C++ projects as libraries.

For projects that adopt multiple build systems, but use only one of them in their CI services, this approach can pose potential risks to the maintenance of build systems as the errors or issues in the unused build systems may not be detected promptly after updating codes. Additionally, using multiple build systems may also cause inconsistency, leading build systems to become out-of-sync, and developers must update all the build systems after each change.

The second reason (35.13\%) for the absence of Bazel usage in CI is that projects can use multiple CI services, with one CI service to build the project and another one to perform management activities on the GitHub repository. For instance, \textit{pytorch/TensorRT} builds the project in CircleCI, while using GitHub Actions to create labels for pull requests or assign developers to issues.

\begin{figure}
\centering
  \includegraphics[width=0.5\textwidth]{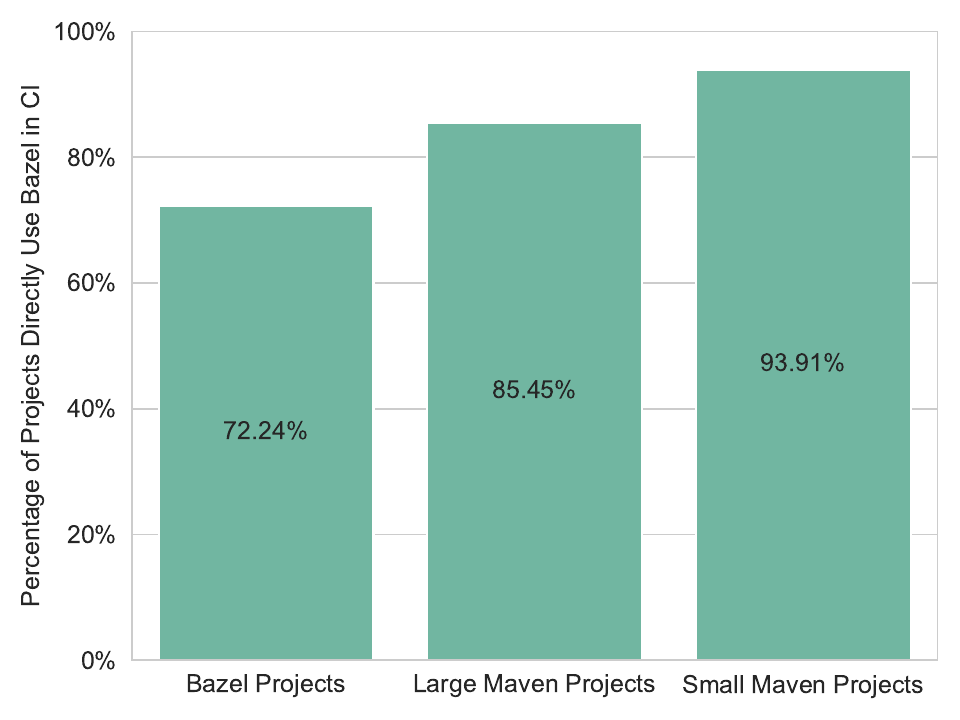}
  \caption{Proportion of direct usage of Bazel/Maven in CI services.}%
  \label{fig:invoke_methods} 
\end{figure}

\begin{table}
  \centering
  \caption{The tools developers use to run Bazel/Maven builds indirectly.}
  \begin{tabularx}{0.6\textwidth}{lXXX}
      \toprule
      Tool & Bazel Projects & Large Maven Projects & Small Maven Projects \\
      \midrule
      Shell Script & 44 & 38 & 11 \\
      \midrule
      Make & 19 & 5 & 2 \\
      \midrule
      Docker & 2 & 2 & 1 \\
      \midrule
      Other & 3 & 3 & 0 \\
      \bottomrule
  \end{tabularx}%
  \label{table:invoker_tools}
\end{table}

\textbf{72.24\% of Bazel projects run Bazel directly in their CI services, whereas approximately 84.45\% of large Maven projects and 93.91\% of small Maven projects run Maven directly.} When using build systems in four studied CI services, developers may run them directly, or use other tools (e.g. Shell Script, Make, Docker) to drive the build systems. Figure~\ref{fig:invoke_methods} illustrates the percentage of projects that directly use build systems in the four studied CI Services.  About 27.76\% of Bazel projects employ other tools to run Bazel, which is higher compared to the other two groups (15.55\% and 6.09\%, respectively). This might suggest that executing Bazel may require a more complicated configuration compared to Maven, leading developers to incorporate configuration tasks into other tools and run Bazel within.

Another possible reason for a higher number of projects running Bazel within other tools may be that projects adopting Bazel have more complicated automation workflows. Table~\ref{table:invoker_tools} outlines the tools used by developers to execute build systems. Shell Script is the most common choice for both build systems. However, 19 projects use Make to run Bazel, which is outside of our initial expectations, as Make is also a build system technology, pre-dating task-based build system technology. Upon examining the Makefiles of these projects, we discovered that most of them utilize Make for managing various automation tasks, including environment setup, and code generation, while using Bazel to build projects is only one of these tasks. In contrast, there is only one project in the large Maven projects group, and none in the small Maven projects group, that utilizes tools like Make for managing their automation tasks.

\begin{figure}
\centering
\subfigure(a){\includegraphics[width=0.45\textwidth]{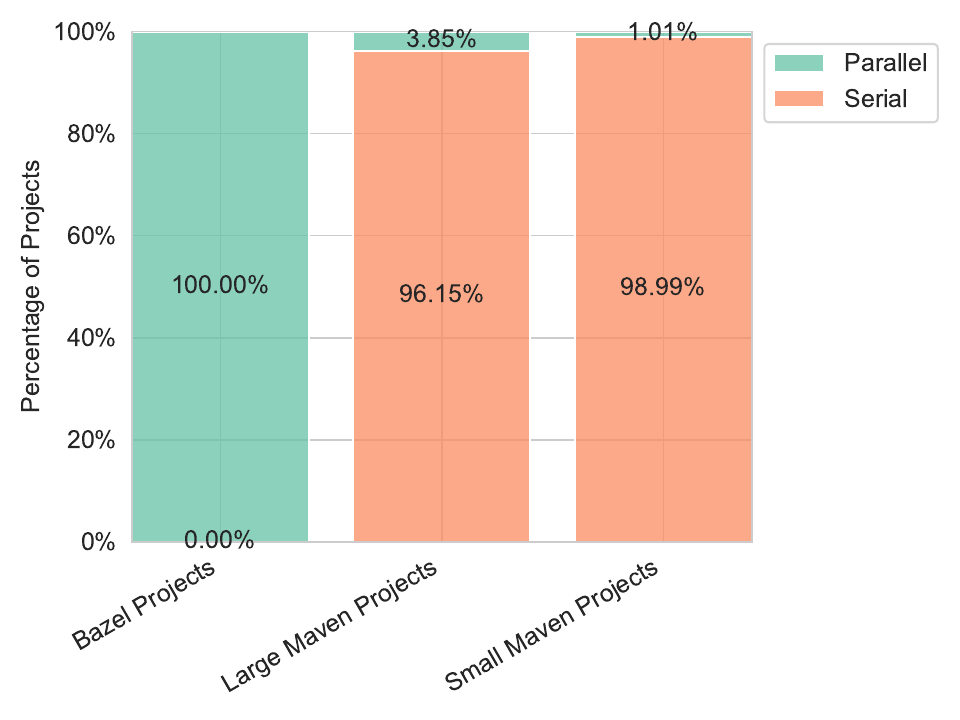}}
\subfigure(b){\includegraphics[width=0.45\textwidth]{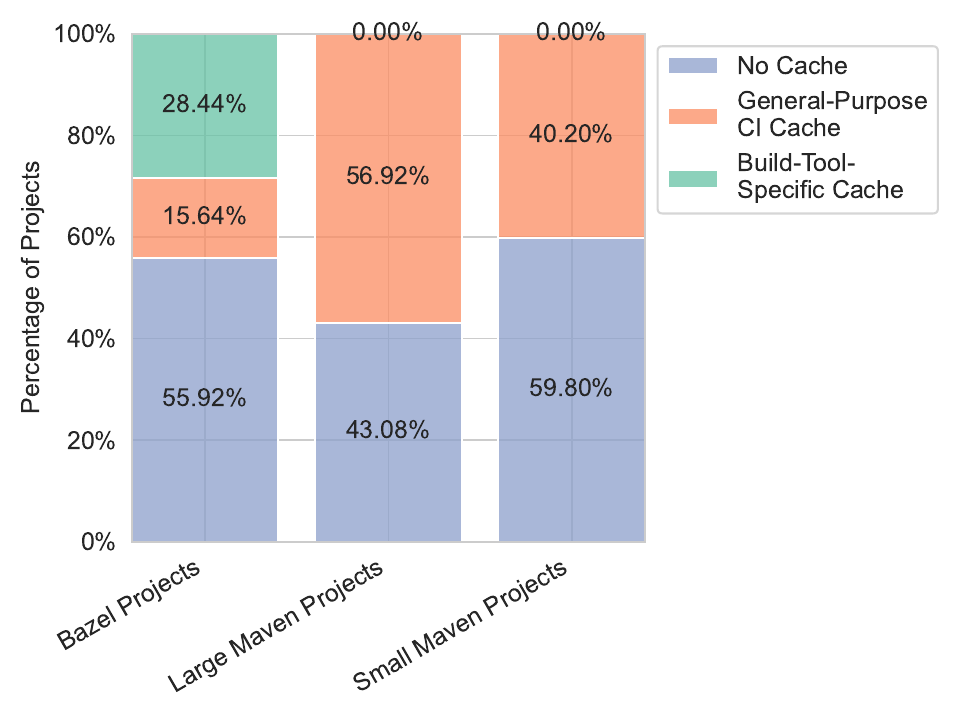}}

\caption{(a) The usage of parallelization in CI services (b) The usage of cache in CI services.}
\label{fig:parallelization_and_cache_usage} 
\end{figure}

\textbf{For projects using build systems in CI services, all Bazel projects use the parallelization feature of Bazel, while there are only 3.85\% of large Maven projects and 1.01\% of small Maven projects that use the parallelization feature of Maven in CI services.} Figure~\ref{fig:parallelization_and_cache_usage} (a) shows the usage of parallelization features of build systems in the four studied CI services. Since parallelization is enabled by default in Bazel, no Bazel project explicitly disables it in CI services. In contrast, for Maven projects, the parallel build functionality needs to be enabled through command-line options or configuration in the \texttt{pom.xml} file. Although the percentage of parallelization usage is higher in the large Maven projects group (3.85\%) compared to the small Maven projects group (1.01\%), both percentages remain low in comparison to Bazel projects (100\%). Our results are consistent with the findings of \cite{testparalleilzation2017}, who studied the usage of test parallelization in open-source Maven projects, showing that only 3.6\% of 468 projects use test parallelization.

However, while parallel builds can potentially reduce build times, they can also introduce challenges such as race conditions caused by unspecified dependencies in the build systems  \citep{bezemer2017empirical}\citep{licker2019detecting}. The widespread adoption of parallelization in Bazel projects does not necessarily suggest that Bazel would be more reliable than Maven when it comes to parallel builds. Another explanation might be that developers may just not know that parallelization is enabled by default in Bazel. RQ2 will analyze parallel build usage in more detail.

\textbf{According to Figure 13 (b), among projects using their build system in the four studied CI services, the usage of caching in CI services for Bazel projects (44.08\%) is lower compared to larger Maven projects (56.92\%) and slightly higher than smaller Maven projects (40.20\%).} These findings surprised us, as Bazel is known for its strong support of incremental builds, and we initially anticipated that Bazel projects would have a higher cache usage in CI services than Maven projects. 

After examining the CI configuration files, we discovered that most Maven projects only cache the project dependencies, whereas Bazel projects cache both dependencies and previous build results. Therefore, even though a high percentage of Maven projects utilize the build system cache in CI services, they do not employ incremental builds and may gain less benefit from cache usage compared to Bazel projects.

Another interesting finding is that although Maven supports Build-Tool-Specific Cache via the Maven Build Cache extension\citep{mavendoc}, Maven projects only use the General-Purpose CI Cache in CI services, while Bazel projects can either use a Build-Tool-Specific Cache or a General-Purpose CI Cache, with higher usage in CI services of the Build-Tool-Specific Cache (28.44\%) compared to the General-Purpose CI Cache (15.64\%). This can be attributed to the Maven Build Cache extension being a newly introduced feature. Therefore, developers may not realize its existence.

In principle, the use of Build-Tool-Specific Cache offers certain advantages and disadvantages. On the positive side, since the cache is not restricted to CI services, developers can share caches not only among CI builds within a single CI service but also between builds across different CI services or between local machine builds and CI service builds. However, developers must also maintain the cache server responsible for storing the caches, which can potentially increase maintenance costs. 

In RQ3, we will further analyze the usage of cache in the CI context. We will compare the performance of the General-Purpose CI and the Build-Tool-Specific Cache, as well as the impact on build performance of caching only dependencies versus caching both dependencies and build results.

\begin{Summary}{}{firstsummary}
For the four studied CI services, 31.23\% of Bazel projects adopt a CI service but do not use Bazel in the CI service. For projects that use Bazel in CI, 27.76\% of them integrate other tools to facilitate Bazel execution in CI services. While all Bazel projects employ parallelization in the four studied CI services, only 3.85\% of large and 1.01 of small Maven projects enable parallelization in CI. Moreover, while the usage of a cache is similar for Bazel (44.08\%), large Maven (56.92\%) and small Maven (40.20\%) projects, most Maven projects only cache the project dependencies, whereas Bazel projects cache both dependencies and previous build results.
\end{Summary}

\subsection{RQ2: What is the impact of Bazel parallelization on the build performance?}

\begin{figure}
\centering
  \includegraphics[width=0.8\textwidth]{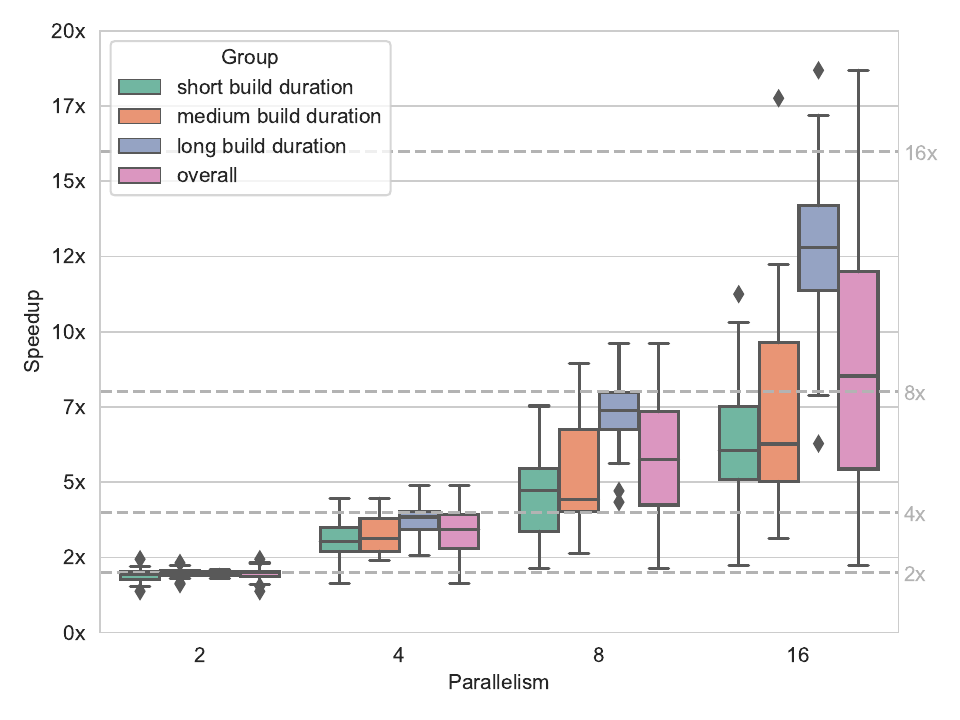}
\caption{Speedup of builds for different parallelism degrees. The projects are categorized into three categories based on their baseline build duration (i.e., the build time with parallelism degree 1)}
\label{fig:parallelization_speedup_by_duration} 
\end{figure}

\textbf{Short, medium, and long-build duration projects show no significant difference in performance improvement at parallelism degrees 2. However, after parallelism degree 2, long-build duration projects demonstrate significantly higher speedups than the other two groups.} Figure~\ref{fig:parallelization_speedup_by_duration} illustrates the speedup in build time for Bazel projects across four parallelism degrees. The speedups are calculated based on their baseline build times, which is the build time at parallelism degree 1. The median speedups for small, medium, and long-build duration projects are 1.92x, 2.00x, and 2.00x at parallelism degree 2. After reaching parallelism degree 4, the speedups for short and medium-build duration projects remain similar, reaching 3.02x and 3.13x at parallelism degree 4, 4.73x and 4.43x at parallelism degree 8, and 6.05x and 6.26x at parallelism degree 16. In contrast, long-duration projects exhibit higher speedups at parallelism degrees 4 (3.84x), 8 (7.36x) and 16 (12.80x).

Table~\ref{table:parallelization_speedup} shows the results of statistical tests conducted between the distributions of build time speedups of short-, medium-, and long-build duration projects at different parallelism degrees. The Kruskal–Wallis test \citep{kruskal1952use} and Dunn's post-hoc test \citep{dunn1964multiple} were employed for analysis. The results indicate that there is indeed no significant difference in speedup across the three groups at parallelism degree 2. However, at parallelism degrees 4, 8, and 16, while there is still no significant difference between short- and medium-build duration projects, long-build duration exhibits significantly higher speedups compared to short- and medium-build duration projects, with large Cliff's Delta \citep{cliff2014ordinal} effect sizes of 0.533 and 0.423 at parallelism degree 4, of 0.772 and 0.591 at parallelism degree 8, and of 0.902 and 0.727 at parallelism degree 16.

\begin{table}
    \caption{Differences in speedups of build time between short, medium, and long-build duration projects. A \colorbox{yellow!50}{yellow cell} represents a medium effect size; a \colorbox{red!50}{red cell} represents a large effect size.}
    \label{table:parallelization_speedup}
      \centering
      \begin{tabularx}{0.95\linewidth}{l *{5}{Y}}
        \toprule
        Comparison\textsuperscript{1} & \multicolumn{4}{c}{Parallelism Degree (p-values of the Dunn's post-hoc test\textsuperscript{2})} \\
        \cmidrule(l){2-5}
        & 2 & 4 & 8 & 16 \\
        \midrule
        S-M & 0.333 & 0.519 & 0.293 & 0.384 \\
        S-L & 0.354 & \cellcolor{red!50}0.004 & \cellcolor{red!50}0.000 & \cellcolor{red!50}0.000 \\
        M-L & 0.794 & \cellcolor{yellow!50}0.03 & \cellcolor{red!50}0.001 & \cellcolor{red!50}0.000 \\
        \midrule
        KW\textsuperscript{3} & 0.230 & 0.005 & 0.000 & 0.000 \\
        \midrule
       \end{tabularx}%
        
    \hspace{1cm}
    
    \begin{tabularx}{\linewidth}{XX}
        \multicolumn{2}{l}{1. S (short-build duration), M (medium-build duration), L (long-build duration).} \\
        \multicolumn{2}{l}{2. Holm is used for adjusting the p-values of Dunn's test.} \\
        \multicolumn{2}{l}{3. Kruskal-Wallis Test.} \\   
    \end{tabularx}
\end{table}

\textbf{The speedups for short- and medium-build duration projects increase slowly after reaching parallelism degree 8. In contrast, the speedups of long-build duration projects exhibit a continuous increase.} Table~\ref{table:parallelization_speedup_between_parallelism_degrees} illustrates the results of the Kruskal–Wallis test and Dunn's post-hoc test applied to the build time speedups between different parallelism degrees within each build duration group. The results show the speedups of short- and medium-build duration projects observed at parallelism degree 8 are not significantly different from those at parallelism 4. In contrast, the long-build duration projects show significant differences between every pair of parallelism degrees, all with large effect sizes.

While, as shown in Table~\ref{table:parallelization_speedup}, short, medium, and long-build duration projects have no significant differences in speedups at parallelism degrees 2, the speedups of long-build duration projects are significantly higher than the other two groups at parallelism degrees 4, 8 and 16. The reason can be two-fold.

First, long-build duration projects have higher costs of communication and synchronization overhead, which offset the performance gain at low parallelism degrees. In Bazel, the build process is divided into multiple actions, and Bazel determines the execution of these actions by analyzing the dependency graph of the projects. Long-build duration projects often have large code bases, leading to a higher number of actions that need to be executed in the build process. As the number of parallel tasks increases, the costs of communication and synchronization overhead become more significant, affecting the efficiency of parallel execution \citep{barney2010introduction}. Additionally, the intricate inter-dependencies among components in large projects can further increase the costs of synchronization. As a result, for long-build duration projects, the gains in performance at lower parallelism degrees may be offset by these costs.

Second, since the potential speedup from parallelization is limited by the serial portions \citep{amdahl1967validity}, short-build duration projects may largely parallelize their parallelizable build parts at parallelism degree 4, while medium-build duration projects do so at 8. Consequently, they experience fewer performance improvements at higher parallelism degrees.

\begin{table}
\centering
\caption{Differences in speedups of build time between different parallelism degrees. A \colorbox{yellow!50}{yellow cell} represents a medium effect size; a \colorbox{red!50}{red cell} represents a large effect size.}
\label{table:parallelization_speedup_between_parallelism_degrees}
\begin{tabularx}{\textwidth}{l *{7}{Y}}
\toprule
Group & \multicolumn{1}{c}{KW\textsuperscript{1}} & \multicolumn{5}{c}{Dunn's Test (p-values) \textsuperscript{2}} \\
\cmidrule(lr){2-2} \cmidrule(l){3-8}
& p & 2-4 & 2-8 & 2-16 & 4-8 & 4-16 & 8-16 \\
\midrule
short build duration & 0.000 & \cellcolor{red!50}0.007 & \cellcolor{red!50}0.000 & \cellcolor{red!50}0.000 & \cellcolor{yellow!50}0.017 & \cellcolor{red!50}0.000 & 0.077 \\
medium build duration & 0.000 & \cellcolor{red!50}0.003 & \cellcolor{red!50}0.000 & \cellcolor{red!50}0.000 & \cellcolor{red!50}0.005 & \cellcolor{red!50}0.000 & 0.092 \\
long build duration & 0.000 & \cellcolor{red!50}0.007 & \cellcolor{red!50}0.000 & \cellcolor{red!50}0.000 & \cellcolor{red!50}0.007 & \cellcolor{red!50}0.000 & \cellcolor{red!50}0.009 \\
\midrule

\multicolumn{8}{l}{1. Kruskal-Wallis Test} \\
\multicolumn{8}{l}{2. Holm is used for adjusting p-values of Dunn's test} \\
\end{tabularx}
\end{table}

\begin{figure}
\centering
\includegraphics[width=0.8\textwidth]{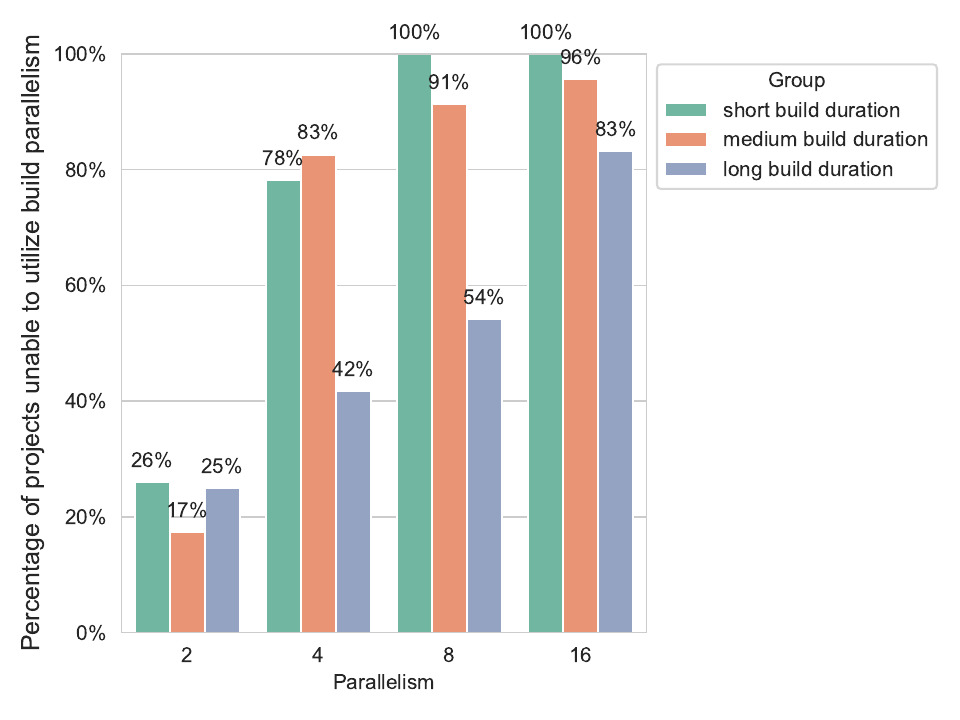}
\caption{Percentage of projects not fully utilizing parallelism.}
\label{fig:parallelization_utilization} 
\end{figure}

\textbf{At parallelism degrees 8 and 16, all short projects and the majority of medium projects (91\% at degree 8 and 96\% at degree 16) are unable to fully exploit parallelism.} Figure~\ref{fig:parallelization_utilization} shows the percentages of projects not leveraging build parallelism within each group. A project does not fully utilize the parallelism if its upper bound of the 95\% confidence interval of the mean speedup is lower than the parallelism degree. At parallelism degree 2, 26\% of the short, 17\% of the medium, and 25\% of the long-build duration projects are not able to fully utilize the parallelism. As the parallelism degree increases, more projects, particularly short and medium-build duration ones, fail to leverage the full potential of parallelism. At parallelism degree 4, this rises to 78\% for short and 83\% for medium-build duration projects, while long-build duration projects stand at 42\%. At degree 8, all short and 91\% of medium-build duration projects fall short in fully utilizing parallelism, compared to 54\% for long-build duration projects. This pattern aligns with prior findings where speedups increase slowly after reaching parallelism degrees 4 for small and 8 for medium-build duration projects.

Furthermore, although the earlier observations indicated continuous performance improvement for long-build duration projects, at parallelism degree 16, there are around 83\% of the projects not able to fully utilize parallelism. 

\begin{table}
\caption{Hierarchical analysis of linear regression models along the three groups of measurements of dependency graph structure and granularity of the compilation units. M1 - granularity-related measures; M2 - Coupling-related measures; M3 - Cohension-related measures.  * p<0.05, ** p<0.01, *** p<0.001. }
    \label{table:dag_analysis}
    \begin{minipage}{\linewidth}
      \centering
      \begin{tabularx}{0.95\linewidth}{l *{5}{Y}}
      \toprule
      M. & MB (Baseline) & M1 & M2 & M3 \\
      \midrule
      BTIME & 0.014 & 0.016 & -0.007 & -0.009 \\
      \midrule
      MNS & & -0.003 & 0.001 & 0.002 \\
      \midrule
      MTD & & & 0.026 & 0.023 \\
      ODS & & & -0.002 & 0.003 \\
      IDS & & & 0.003 & -0.002 \\
      \midrule
      WCC & & & & 0.090 \\
      CC & & & & 0.003 \\
      ASPL & & & & 0.003 \\
      \midrule
      R\textsuperscript{2} & 0.012 & 0.014 & 0.058 & 0.063 \\
      \bottomrule
      \end{tabularx}%
    \end{minipage}%
    \hspace{1cm}
    
    \begin{tabularx}{\linewidth}{X}
        \textbf{(a)} Linear regression coefficients and p-values at parallelism degree 2.
    \end{tabularx}
    \hspace{2cm}
    \begin{minipage}{\linewidth}
      \centering
      \begin{tabularx}{0.95\linewidth}{l *{5}{Y}}
      \toprule
      M. & MB (Baseline) & M1 & M2 & M3 \\
      \midrule
      BTIME & \textbf{0.186 **} & \textbf{0.218 **} & \textbf{0.210 *} & \textbf{0.188 *} \\
      \midrule
      MNS & & -0.046 & -0.042 & -0.046 \\
      \midrule
      MTD & & & 0.025 & 0.009 \\
      ODS & & & -0.014 & -0.020 \\
      IDS & & & -0.003 & -0.004 \\
      \midrule
      WCC & & & & 0.493 \\
      CC & & & & 0.018 \\
      ASPL & & & & -0.289 \\
      \midrule
      R\textsuperscript{2} & 0.129 & 0.147 & 0.184 & 0.207 \\
      \bottomrule
      \end{tabularx}%
    \end{minipage}%
    \hspace{1cm}
    
    \begin{tabularx}{\linewidth}{X}
        \textbf{(b)} Linear regression coefficients and p-values at parallelism degree 4.
    \end{tabularx}
    \hspace{2cm}
    \begin{minipage}{\linewidth}
      \centering
      \begin{tabularx}{0.95\linewidth}{l *{5}{Y}}
      \toprule
      M. & MB (Baseline) & M1 & M2 & M3 \\
      \midrule
      BTIME & \textbf{0.734 ***} & \textbf{0.798 ***} & \textbf{0.824 ***} & \textbf{0.824 ***} \\
      \midrule
      MNS & & -0.089 & -0.087 & -0.078 \\
      \midrule
      MTD & & & 0.003 & 0.021 \\
      ODS & & & -0.060 & -0.057 \\
      IDS & & & 0.003 & 0.002 \\
      \midrule
      WCC & & & & 0.485 \\
      CC & & & & 0.003 \\
      ASPL & & & & -0.219 \\
      \midrule
      R\textsuperscript{2} & 0.284 & 0.294 & 0.325 & 0.326 \\
      \bottomrule
      \end{tabularx}%
    \end{minipage}%
    \hspace{1cm}
    
    \begin{tabularx}{\linewidth}{X}
        \textbf{(c)} Linear regression coefficients and p-values at parallelism degree 8.
    \end{tabularx}
    \hspace{2cm}
    \begin{minipage}{\linewidth}
      \centering
      \begin{tabularx}{0.95\linewidth}{l *{5}{Y}}
      \toprule
      M. & MB (Baseline) & M1 & M2 & M3 \\
      \midrule
      BTIME & \textbf{1.890 ***} & \textbf{2.253 ***} & \textbf{2.518 ***} & \textbf{2.544 ****} \\
      \midrule
      MNS & & \textbf{-0.509 **} & \textbf{-0.568 **} & \textbf{-0.514 **} \\
      \midrule
      MTD & & & -0.195 & -0.076 \\
      ODS & & & -0.020 & -0.023 \\
      IDS & & & -0.050 & -0.022 \\
      \midrule
      WCC & & & & 1.982 \\
      CC & & & & -0.179 \\
      ASPL & & & & -0.720 \\
      \midrule
      R\textsuperscript{2} & 0.416 & 0.488 & 0.526 & 0.536 \\
      \bottomrule
      \end{tabularx}%
    \end{minipage}%
    \hspace{1cm}
    \begin{tabularx}{\linewidth}{X}
        \textbf{(d)} Linear regression coefficients and p-values at parallelism degree 16.
    \end{tabularx}
    \hspace{2cm}
\end{table}

\textbf{The baseline build time is significantly correlated with build time speedups at parallel degrees 4, 8, and 16, while the average size of the compilation unit shows a significant correlation specifically at a parallelism degree of 16.} The results of the linear regression analysis at these four parallelism degrees are presented in Table~\ref{table:dag_analysis}. We started the linear regression analysis using the baseline build time as the baseline model (Model MB) since we already found that there is a relationship between baseline build time and build speedups (dependent variable). Model M1 incorporated granularity-related measures, Model M2 included coupling-related measures, and Model M3 introduced cohesion-related measures.  As shown in Table~\ref{table:dag_analysis}, at parallelism degree 2, no measurement shows a significant relationship with the speedups of the build, while at parallelism degrees 4, 8, and 16, the baseline build time shows a significant correlation with the speedups with positive coefficients, which is consistent with our previous findings. Surprisingly, measures related to the dependency graph do not show any significant relationship with speedups at all four parallelism degrees. Notably, the size of the compilation unit only displays a significant correlation at a parallelism degree of 16 with negative coefficients, suggesting the smaller the compilation units, the higher the possibility the project gains better performance at parallelism degree 16.

\begin{Summary}{}{secondsummary}
Given that most CI services default to having only 2 cores, there is potential to make the Bazel build performance better by increasing the number of cores. However, for short- and medium-build duration projects, a parallelism degree beyond 4 might not be useful. While developers can still benefit from the parallelism, they should consider if the performance gain is worth the extra hardware costs. On the other hand, long-build duration projects experience continuous performance improvement as parallelism increases, but fewer projects can fully leverage it. Moreover, we found no significant correlation between the dependency graph structure and the build performance across all four parallelism degrees, while the average size of compilation units shows a significant correlation at parallelism degree 16.
\end{Summary}

\subsection{RQ3: What is the impact of Bazel incremental build (cache) functionality on the build performance?}

\begin{figure}
    \centering
    \includegraphics[width=0.8\textwidth]{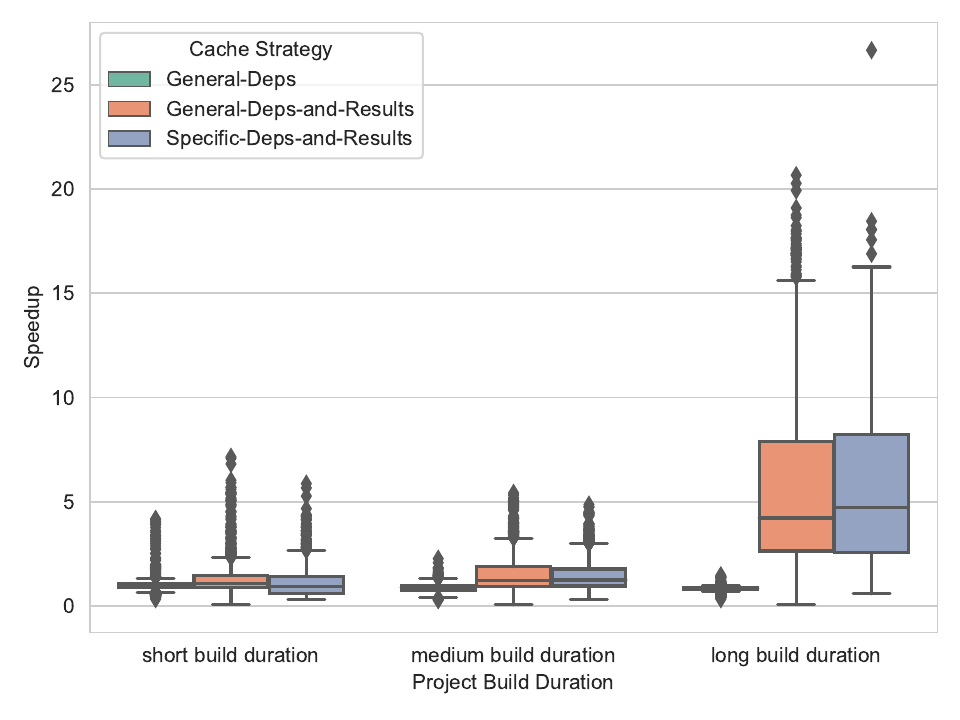}
    \caption{The speedup of build time with different cache strategies}
    \label{fig:cache_speedup}
\end{figure}

\textbf{When only caching build dependencies in CI builds, 8.70\% of medium and 4.17\% of long-build duration projects exhibit improved build performance compared to clean builds, while 52.17\% of short-build duration projects show better performance.} Figure~\ref{fig:cache_speedup} shows the speedups achieved by projects using different cache strategies. The projects are divided into three groups based on their median build time of clean builds. The median speedups of short, medium, and long-build duration projects with the \textit{General-Deps} strategy are 1.00x, 0.88x, and 0.85x, respectively. We calculated the 95\% mean confidence interval of speedups of the \textit{General-Deps} strategy experiments. The results reveal that only a small portion of medium (8.70\%) and long-build (4.17\%) projects exhibit better performance (i.e., the lower bound of the mean confidence interval of speeds is higher than 1) than the baseline clean builds. Although short-build duration shows a higher (52.17\%) percentage of projects that demonstrate improved performance than clean builds with the \textit{General-Deps} strategy, this figure still falls just around half of the total projects in this category.

Caching dependencies in CI build is a common practice in open-source projects to reduce the build time. However, the results indicate that caching dependencies alone may not help the build performance too much especially for projects with longer build duration. Although in practice, CI services can use techniques like compression or deploying the cache storage close to machines running CI builds to reduce the time of downloading and uploading dependencies caches in builds, our results are still meaningful for long-build duration projects. Because long-build duration projects usually have only a small portion of build time used to download dependencies, using dependency cache alone may not lead to a substantial reduction in overall build time for such projects. Our findings are reinforced by the data shown in Table~\ref{table:cache_speedup} (b), where the speedups of medium and long-build duration projects with \textit{General-Deps} strategy are actually significantly lower than short-duration projects with a medium (0.407) and a large (0.572) effect size, respectively.

\begin{table}
    \caption{Differences in speedups between (a) cache strategies within the build duration group and (b) build duration groups under the same cache strategy. A \colorbox{gray!50}{grey cell} represents a negligible effect size; a \colorbox{blue!30}{blue cell} represents a small effect size; a \colorbox{yellow!50}{yellow cell} represents a medium effect size; a \colorbox{red!50}{red cell} represents a large effect size.}
    \label{table:cache_speedup}
    \begin{minipage}{0.5\linewidth}
      \centering
      \begin{tabularx}{0.95\linewidth}{l *{4}{Y}}
      \toprule
      Comp.\textsuperscript{1}\textsuperscript{2} & \multicolumn{3}{c}{Group\textsuperscript{3}} \\
      \cmidrule(l){2-4}
      & S & M & L \\
      \midrule
      GD-GDB & \cellcolor{blue!30}0.000 & \cellcolor{red!50}0.000 & \cellcolor{red!50}0.000 \\
      GD-SDB & \cellcolor{gray!50} 0.002 & \cellcolor{red!50}0.000 & \cellcolor{red!50}0.000 \\
      GDB-SDB & \cellcolor{blue!30}0.000 & 1.000 & 0.08 \\
      \midrule
      KW\textsuperscript{4} & 0.000 & 0.000 & 0.000 \\
      \end{tabularx}%
    \end{minipage}%
    \begin{minipage}{0.5\linewidth}
      \centering
      \begin{tabularx}{0.95\linewidth}{l *{4}{Y}}
      \toprule
      Comp. & \multicolumn{3}{c}{Cache Strategy\textsuperscript{3}} \\
      \cmidrule(l){2-4}
      & GD & GDB & SDB \\
      \midrule
      S-M & \cellcolor{yellow!50}0.000 & \cellcolor{blue!30}0.000 & \cellcolor{yellow!50}0.000 \\
      S-L & \cellcolor{red!50}0.000 & \cellcolor{red!50}0.000 & \cellcolor{red!50}0.000 \\
      M-L & \cellcolor{gray!50}0.000 & \cellcolor{red!50}0.000 & \cellcolor{red!50}0.000 \\
      \midrule
      KW & 0.000 & 0.000 & 0.000 \\
      \end{tabularx}%
    \end{minipage}%
    \hspace{1cm}
    
    \begin{tabularx}{\linewidth}{XX}
        \textbf{(a)} Differences between cache strategies. & \textbf{(b)} Differences between groups. \\ 
        \addlinespace[0.5cm]
        \multicolumn{2}{l}{1. GD (\textit{General-Deps}), GDB (\textit{General-Deps-and-Results}), SDB (\textit{Specific-Deps-and-Results}).} \\
        \multicolumn{2}{l}{2. S (Short-build duration), M (Medium-build duration), L (Long-build duration).} \\
        \multicolumn{2}{l}{3. P-values of Dunn's post-hoc test. Holm is used for adjusting the p-values.} \\
        \multicolumn{2}{l}{4. Kruskal-Wallis Test.} \\   
    \end{tabularx}
\end{table}

\textbf{Caching build results significantly reduces the build time for medium and long-build duration projects, while it is less effective for short-build duration projects.} The median speedups with \textit{General-Deps-and-Results} and \textit{Specific-Deps-and-Results} strategies for long-build duration projects are 4.22x and 4.71x, respectively. As illustrated in Table~\ref{table:cache_speedup} (a), long-build duration projects have significantly higher speedups using either \textit{General-Deps-and-Results} (with a large effect size of 0.910) and \textit{Specific-Deps-and-Results} (with a large effect size of 0.954) in builds than when using the \textit{General-Deps} strategy. Similar trends are observed for medium-build duration projects, with median speedups of 1.21x and 1.25x, along with large effect sizes of 0.582 and 0.633, respectively.

Conversely, for projects with short build duration, the difference in speedups between the \textit{Specific-Deps-and-Results} and \textit{General-Deps} strategies show only a negligible difference (effect size: 0.071), and the difference between \textit{General-Deps-and-Results} and \textit{General-Deps} strategies is also small (effect size: 0.167). One possible explanation of the lower performance improvement for short-build duration projects may be that the time spent on uploading and downloading caches could outweigh the time saved by the incremental build, given their already short build times. 

Long-build duration projects, on the other hand, particularly benefit from caching build results during builds. As indicated in Table~\ref{table:cache_speedup} (b), long-build duration projects achieve significantly higher speedups with large effect sizes than both medium- and short-build duration with either the \textit{General-Deps-and-Results} or \textit{Specific-Deps-and-Results} strategy.

\textbf{As shown in Table~\ref{table:cache_speedup} (a), \textit{Specific-Deps-and-Results} shows no significant difference in speedups for medium and long build-duration projects and lower speedups with a small (0.179) effect size for short-build duration projects, compared to the \textit{General-Deps-and-Results} strategy.} Our results indicate that there is no or only limited difference between the General-Purpose CI Cache and Build-Tool-Specific Cache in terms of build performance. Considering the similar performance between them, and the lower maintenance effort required for General-Purpose CI Cache, it seems General-Purpose CI Cache may be a better choice for developers. However, using a General-Purpose CI Cache also brings security risks if it is not used well. The Build-Tool-Specific Cache is managed by the build system and the build system knows which files should be cached, but, for General-Purpose CI Cache, developers usually specify the file system paths in the CI configuration file, and the CI services cache all the files in these paths. Consequently, there may be some files containing sensitive data under that path that are accidentally uploaded to the cache and, therefore can be accessed by others. So, developers should be careful when using the General-Purpose CI Cache to ensure not caching any sensitive files.

\begin{figure}
  \centering
  \includegraphics[width=0.6\textwidth]{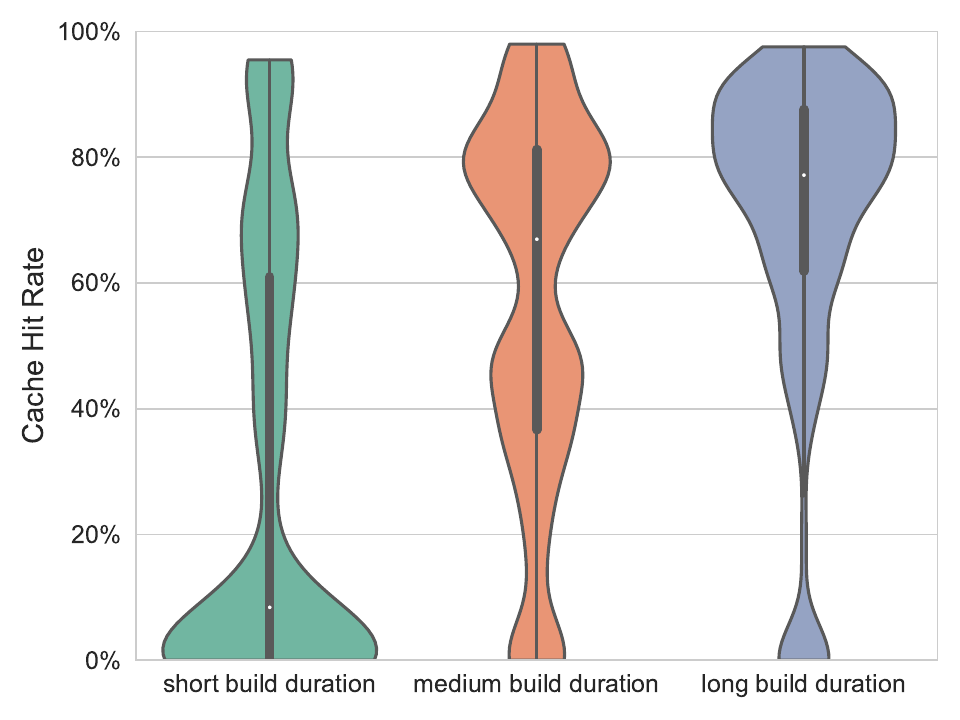}
\caption{The cache hit rates of Bazel projects in the experiments}
\label{fig:cache_hit_ratio} 
\end{figure}

\textbf{The median cache hit rate for short, medium, and long-build duration projects are 8.42\%, 66.96\%, and 77.15\%, respectively.} Figure~\ref{fig:cache_hit_ratio} illustrates the cache hit rate of Bazel projects in the experiments. The medium and long build duration projects have significantly higher cache hit rates than the short build duration projects, with effect sizes 0.446 (medium) and 0.564 (large), respectively. 

Additionally, the long build duration projects also demonstrate significantly higher cache hit rates than the medium build duration projects with a small effect size of 0.241. Since short-build duration projects usually have smaller code bases, changes with the same size in short-build duration projects could have a higher impact on cache hit rates than in medium and long-build duration projects.

\begin{Summary}{}{thridsummary}
Caching dependencies in CI builds is a common practice to reduce the build time, yet employing only such caching may not always be effective, especially for medium and long-build duration projects. On the other hand, using incremental builds significantly improves the build performance in CI for medium and long-build duration projects, with only a minor impact on build times for short-build duration projects. Additionally, both General-Purpose CI Cache and Build-Tool-Specific Cache show similar speed improvements when used for incremental builds. However, developers should keep in mind that with General-Purpose CI Cache, the developers themselves are deciding which files to save to or restore from caches, which can bring potential security vulnerabilities since sensitive data may be inadvertently written to caches and accessed by others. In contrast, with a Build-Tool-Specific Cache, where Bazel manages caches and knows which files should be cached, such security concerns are less likely to occur.
\end{Summary}

\section{Discussion}
\label{sec:discussion}
In this section, we discuss the implications of our results to practitioners and researchers.

\textit{For developers who use Bazel in their projects.} Our findings (RQ2 and RQ3) highlight the substantial improvement in build performance that developers can gain through Bazel's parallel and incremental build features. Nonetheless, our results reveal that this potential can be underutilized, especially in the CI context, as we find that developers may not use Bazel at all in CI builds or not use Bazel's incremental build to speed up CI builds (RQ1). 

Given that the default number of cores for most CI services is 2, for projects employing Bazel's parallel and incremental builds in CI, there may still be potential for developers to speed up CI builds (RQ2). Our results show that developers can significantly improve the build performance by increasing the parallelism from 2 to higher degrees (RQ2). Furthermore, our findings indicate that the structure of the dependency graph does not have any significant impact on the parallelization efficiency (RQ2). However, there is a significant correlation between the granularity of dependencies and build speedups at a parallelism degree of 16 (RQ2). For developers, on the one hand, improving software structure may enhance maintainability but does not necessarily improve the benefits of parallelization in the build process. On the other hand, reducing the granularity of artifacts proves beneficial for parallelization efficiency at a parallelism degree of 16. Considering that many projects do not use caching in CI or only cache dependencies in CI (RQ1), our results show that developers are able to significantly speed up the build by applying incremental builds (RQ3).

However, while parallel and incremental builds of Bazel significantly improve the build performance, developers should keep in mind that these features also have limitations. Increasing the parallelism in CI builds reduces the build time, but after reaching a certain parallelism degree, the improvements slow down (RQ2). Applying incremental build can speed up the builds, however, it may not be effective for projects with short build duration (RQ3). 

Furthermore, the maintenance costs of Bazel can be higher in comparison to traditional build systems such as Maven. As shown in Table~\ref{table:divided_bazel_projects}, we applied the same threshold of 731.25 commits previously used (in section 3.1) for dividing Maven projects to now also divide Bazel projects into large and small Bazel project groups, comprising 223 and 160 projects, respectively. While large Maven projects exhibit lower direct build tool usage in CI services compared to the small Maven project group, which can be attributed to the complexity of workflows in larger projects, even small Bazel projects exhibit lower direct build tool usage in CI compared to large Maven projects, suggesting a higher complexity of configuration and executing Bazel in CI. This confirms the finding of \cite{alfadel2024icse}'s study that developers often face challenges when integrating Bazel with other platforms and tools. In addition, both Bazel groups show lower build tool usage in CI, primarily because developers adopt multiple build systems and use other build systems for CI services (RQ1). These practices are partly responsible for the abandonment of Bazel by organizations due to the maintenance costs of keeping multiple build systems in sync \citep{alfadel2024icse}.

Therefore, developers need to carefully weigh the performance benefits against hardware resource costs and maintenance efforts when using parallel and incremental builds in their CI builds.

\begin{table}[]
    \caption{Results of RQ1 with Bazel projects being divided into two groups using the same threshold previously applied for dividing Maven projects.}
    \label{table:divided_bazel_projects}
    \begin{minipage}{\linewidth}
    \centering
    \begin{tabularx}{\linewidth}{l *{6}{Y}}
        \toprule
        \multirow{2}{*}{Dataset} & \multicolumn{3}{c}{Bazel} & \multicolumn{3}{c}{Maven} \\
        & Overall & Large & Small & Overall & Large & Small \\
        \midrule
        RQ1.1 & 68.77\% & 64.98\% & 74.81\% & 91.85\% & 90.52\% & 93.70\% \\
        RQ1.2 & 72.24\% & 70.97\% & 74.77\% & 88.93\% & 85.45\% & 93.91\% \\
        RQ1.3 & 100\% & 100\% & 100\% & 2.61\% & 3.85\% & 1.01\% \\
        RQ1.4 & 44.08\% & 42.74\% & 45.34\% & 49.67\% & 56.92\% & 40.20\% \\
        \bottomrule
        \addlinespace[0.3cm]
        \multicolumn{7}{l}{\makecell[l]{RQ1.1 - The percentage of projects adopting a CI service and using the Bazel/Maven \\ build system in their CI service.}} \\
        \multicolumn{7}{l}{\makecell[l]{RQ1.2 - The percentage of projects directly using the Bazel/Maven build system in CI.}} \\
        \multicolumn{7}{l}{\makecell[l]{RQ1.3 - The percentage of projects using the parallelization feature in CI.}} \\
        \multicolumn{7}{l}{\makecell[l]{RQ1.4 - The percentage of projects using the cache feature in CI.}} \\
    \end{tabularx}
    \end{minipage}
\end{table}

\textit{For build technology developers.} Our study emphasizes the importance of guiding users in the use of build technology features. As a previous study pointed out \citep{mokhov2018build}, while build systems are a fundamental part of software development, they receive limited attention from developers. Our results reveal a notable disparity between available features and their utilization (RQ1). Considering the benefits these features bring to build performance (RQ2 and RQ3), developers should pay more attention to how to guide build system users to use build technology features and simplify their configurations. Default enabling of some features, if possible, may also be a solution. For example, parallel build is by default enabled in Bazel, and all Bazel projects use it in CI, while Maven users need to enable parallel build by command-line options, and only 2.27\% of large Maven and 0.55\% of small Maven projects use parallel build in CI. Although this disparity can be explained by the difference between Bazel and Maven, considering the lack of attention from developers to build technology \citep{mokhov2018build}, a lot of them may just not realize the existence of such features and their benefits.

Furthermore, build technology developers should provide users with best practices on how to effectively harness these features. Our findings show that longer-build duration projects typically benefit more from parallel and incremental builds (RQ2 and RQ2) than shorter-build duration projects, or that caching dependencies alone is not effective in reducing build time for longer-build duration projects (RQ3), could be guidance aiding users in understanding the best application scenarios for each feature, thus facilitating more efficient usage.

\textit{For researchers.} While traditional task-based build tools have been extensively studied, modern artifact-based build tools like Bazel have received less attention. Our study investigates Bazel's utilization and the benefits of its parallel and incremental builds within the CI context. However, as projects are organized differently with modern build systems, further exploration into how the adoption of these systems impacts software development is required.

Investigating distributed builds (an important feature of Bazel) may also be an area researchers need to look at. Given the prevalence of cloud computing and the trend of cloud-based CI workflows \citep{garg2019automated}, investigating distributed build's performance, correctness, and security in cloud environments could yield valuable insights.

Lastly, the increasingly frequent rate at which software is delivered driven by industry practices like agile development and microservices \citep{railic2021architecting}, leads to more frequent builds in CI. While our study examines the CI usage and usefulness of Bazel, we notice that there is still a lack of attention to investigating build system usage in the CI context. Hence, we believe that further empirical studies are necessary to better understand the usage of build systems in the CI context.

\section{Threats to Validity}
\label{sec:threats}
\subsection{Construct Validity}
We used SourceGraph's search API to identify projects that use Bazel. However, SourceGraph may not be able to index all GitHub repositories, which can impact the accuracy of our dataset. To detect the usage of Bazel in GitHub repositories, we relied on the presence of specific Bazel-related files. But the file name used is very common (i.e. BUILD), which is also used by other build systems like Buck and Pants as their default name of configuration files. We mitigated this issue by using Linguist, a tool open-sourced by GitHub, to identify the programming language of the build file and only kept the projects whose build file is written by Starlark. Since this tool is not always accurate, after employing this tool, we manually examined each project to ensure we only included Bazel projects in the dataset.

Moreover, we used the presence of CI configuration files as evidence that the projects are using the CI services. But, this is not necessarily true, since projects may drop the use of a CI service but still keep the CI configuration files. This is another threat that might impact the validity of the research. We looked at the CI configuration files to analyze the build commands of Bazel and Maven. Since we only looked at the default locations for these configuration files, we may miss some configuration files as the locations can be configured through their web interface. This is a well-known problem that has been reported by other researchers \citep{bird2009promises} \citep{vasilescu2015quality}.

In addition, while we removed projects with less than 100 commits and stars to exclude trivial projects, there is still a possibility that some toy projects (e.g., tutorial, example projects) were included in the dataset. Therefore, we conducted a manual inspection of the projects within the dataset. In RQ1, where our analysis focused exclusively on projects adopting GitHub Actions, Travis CI, CircleCI, or Buildkite as their CI services, we found 0 Bazel (0\%), 2 large Maven (0.7\%), and 12 small Maven (5.66\%) projects that are toy projects. In RQ2 and RQ3, our analysis focused on 70 buildable Bazel projects, none of which were toy projects. Given the low number of toy projects, we believe that our results remain valid.

\subsection{Internal Validity}
In RQ1, we checked the entire population of Bazel projects in the datasets and selected CI-using projects from them. We employed the same approach to sample Maven projects in the datasets for fairness. However, this approach resulted in an unequal size of groups between Bazel (289 projects), large Maven (282 projects), and small Maven (201 projects) project groups. To mitigate this potential bias, we use percentages instead of absolute numbers to compare the results of each group.

We developed Python scripts to collect projects and analyze their CI configuration files in RQ1. However, the scripts may not be able to capture all the data we need in the dataset, which can affect the accuracy of the results. For example, we followed the CI services' specification to extract the shell commands executed in the workflow and identify the Bazel and Maven-related shell commands from them. However, developers may not directly execute the Bazel in the shell commands but instead run other tools and execute Bazel within. To mitigate this threat, we also analyzed shell script files and Makefiles of the projects, if the shell script files or Make are used in the shell commands. Some projects have complex workflows, for instance, \textit{google/gvisor} runs a Docker container in its CI workflow, and inside the container, it uses Make to execute Bazel to build the project. Our scripts are not able to identify the build system-related shell commands in such cases. Therefore, for projects in which we did not identify any Maven or Bazel-related shell commands, we manually examined them to extract the commands.

Furthermore, in RQ3, we employed FUSE to mount the object storage buckets onto the machine running the experiments to simulate the General-Purpose CI Cache. In practice, CI providers use different techniques, like compression or deploying object storage in the same region as machines, to optimize the performance of the cache. To mitigate this issue, all the machines and object storage buckets are deployed onto the same region. However, we did not employ compression to optimize the General-Purpose CI Cache, which could result in lower performance and impact the validity of the study.

\subsection{External Validity}
In RQ1, we only analyzed the CI configuration files of four CI services (GitHub Actions, CircleCI, Travis CI, and Buildkite). The results may differ for projects using other CI/CD services. However, since the first three CI services are the three most popular CI/CD services on GitHub \citep{golzadeh2022rise}, and the last one is popular among Bazel projects, we believe the results are representative. 

In RQ2 and RQ3, we evaluated the parallel build and incremental build for Bazel open-source projects. Since Bazel is an artifact-based build system, the results may not be applicable to task-based build systems. However, for artifact-based build systems, since most of them are the derivatives of Blaze (the internal version of Bazel at Google), the results can be generalized to them.

Moreover, RQ2 and RQ3 focus only on compilation activities. We believe our findings to be also helpful in guiding developers with test execution, given that the mechanisms for parallel build and incremental build execution both apply for compilation and for test execution. However, the effectiveness of these performance optimizations on testing still needs to be investigated. We intend to extend our research to include test execution in future studies.

\section{Conclusion}
\label{sec:conclusion}
In this paper, we investigate how developers employ Bazel within their CI builds to build projects, along with exploring the benefits and limitations associated with Bazel's parallel and incremental builds in the CI context.

Our findings show that Bazel's parallel and incremental build features have substantial potential for significantly reducing build times in CI. However, these features are not always harnessed to their full extent by developers, as these features may not be used or are underutilized by developers in CI builds. Our study also reveals the limitations of parallel and incremental builds of Bazel. We observe that as the parallelism degree increases, fewer projects, particularly for short and medium-build duration projects, can fully exploit the parallelism. Furthermore, when applying incremental build features in CI, we found that long-build duration projects achieve significantly higher speedups than short and medium-build duration projects, while the speedups of the medium group are significantly higher than the short group. 

In future work, as anecdotal evidence shows that tests can benefit more from Bazel than builds \footnote{https://blog.mozilla.org/nfroyd/2019/11/01/evaluating-bazel-for-building-firefox-part-2/}, we intend to conduct a study on Bazel's testing in open-source projects to understand its maintenance and performance. Since in this study, we only evaluated the performance of Bazel's parallel and incremental builds, therefore, we also plan to investigate the correctness and reproducibility of Bazel builds to empirically validate the reliability claims of Bazel.

\section{Data Availability Statement}

The datasets have been made publicly available on GitHub for replication purposes at this link: https://github.com/SAILResearch/replication-23-shenyu-bazel\_usage

\section{Conflicts of Interest}

The authors declare that they have no conflicts of interest.



\def\UrlBreaks{\do\/\do-\do.} 
\bibliographystyle{spmpscinat}

\bibliography{references}   

%
%

\end{document}